\documentclass[psfig]{mn2e}
\usepackage{epsfig}

\newcommand{\ratpol}{($\frac{\Delta \lambda(\rm Pol)}{\Delta \lambda(I)}$)}

\newcommand{\ang}{$\rm \AA$}

\newcommand{\msun}{M$_{\odot}$}

\newcommand{\degree}{$^{\rm o}$}

\newcommand{\ea}{{et al.}}

\newcommand{\ha}{H$\alpha$}

\newcommand{\bb}{\bibitem[]{bla}}

\def\kms{\,km~s$^{-1}$}      % note leading thinspace

\def\lesssim{\mathrel{\hbox{\rlap{\hbox{\lower4pt\hbox{$\sim$}}}\hbox{$<$}}}}
\def\gtrsim{\mathrel{\hbox{\rlap{\hbox{\lower4pt\hbox{$\sim$}}}\hbox{$>$}}}}

\def\arcsec{\hbox{$^{\prime\prime}$}}

\def\ion#1#2{#1$\;${\small\rm\@Roman{#2}}\relax}
 
\newbox\grsign \setbox\grsign=\hbox{$>$} \newdimen\grdimen 
\grdimen=\ht\grsign
\newbox\simlessbox \newbox\simgreatbox
\setbox\simgreatbox=\hbox{\raise.5ex\hbox{$>$}\llap
     {\lower.5ex\hbox{$\sim$}}}\ht1=\grdimen\dp1=0pt
\setbox\simlessbox=\hbox{\raise.5ex\hbox{$<$}\llap
     {\lower.5ex\hbox{$\sim$}}}\ht2=\grdimen\dp2=0pt

\makeatletter 
\renewcommand\@biblabel[1]{}     % Arabic numbers, 

\makeatother

\begin{document}

\title[Probing the circumstellar structure of Herbig Ae/Be stars]
{Probing the circumstellar structure of Herbig Ae/Be stars}

\author[Jorick S. Vink \ea ]
{Jorick S. Vink$^1$, Janet E. Drew$^1$, Tim J. Harries$^2$, Ren\'e D. Oudmaijer$^3$\\
$^1$Imperial College of Science, Technology and Medicine,
Blackett Laboratory, Prince Consort Road, London, SW7 2BZ, UK\\   
$^2$ School of Physics, University of Exeter, Stocker Road, Exeter EX4 4QL, UK\\
$^3$ The Department of Physics and Astronomy, E C Stoner Building, Leeds, LS2 9JT, UK\\ 
}

\date{received,  accepted}

\maketitle
\begin{abstract}

We present \ha\ spectropolarimetry observations of a sample of 23
Herbig Ae/Be stars. A change in the linear polarisation across 
\ha\ is detected in a large fraction of the objects, 
which indicates that the regions around Herbig stars are 
flattened (disc-like) on small scales. 
A second outcome of our study is that the spectropolarimetric 
signatures for the Ae stars differ from those of the Herbig Be 
stars, with characteristics changing from depolarisation across \ha\ 
in the Herbig Be stars, to line polarisations in the Ae group. 
The frequency of depolarisations detected in the Herbig Be stars (7/12) is 
particularly interesting as, by analogy to classical Be stars, it 
may be the best evidence to date that the higher mass Herbig stars 
are surrounded by flattened structures.
For the Herbig Ae stars, 9 out of 11 show a line polarisation effect 
that can be understood in terms of a compact \ha\ emission that is 
itself polarised by a rotating disc-like circumstellar medium.
The spectropolarimetric difference between the Herbig Be 
and Ae stars may be the first indication that there 
is a transition in the Hertzsprung-Russell Diagram from magnetic 
accretion at spectral type A to disc accretion at spectral type B. 
Alternatively, the interior polarised line emission apparent
in the Ae stars may be masked in the Herbig Be stars due to their 
higher levels of \ha\ emission.

\end{abstract}

\begin{keywords}

stars: formation       --
stars: pre-main sequence    --
stars: Herbig Ae/Be         --
circumstellar matter --
techniques: polarimetric
 \end{keywords}

\section{Introduction}

Although we seem to understand reasonably well how low
mass stars form by cloud collapse and disc accretion, the 
question of whether high mass stars form in the same 
way remains completely open.
On the one hand, theoretical efforts are being undertaken 
to realise the formation of high mass stars through disc 
accretion (e.g. Behrend \& Maeder 2001). 
On the other hand, others perform calculations for the formation 
of high mass stars by stellar collisions and mergers in a dense  
cluster environment (e.g. Bonnell, Bate \& Zinnecker 1998). 
These collision scenarios are invoked because high mass star formation 
through disc accretion encounters the problem that 
radiation pressure forces on gas and dust may be able to reverse 
the infall, preventing the formation of stars with masses larger than 
10 \msun (York \& Kruegel 1977).

In addressing the question as to whether accretion discs around 
high mass stars are present, the intermediate mass 
(2 -- 15 \msun) pre-main sequence Herbig Ae/Be stars (Herbig 1960) 
play a crucial role, as these objects are the only higher mass 
pre-main sequence stars that are visible at optical and infrared 
wavelengths.
Yet, whether or not Herbig Ae/Be stars are embedded in
accretion discs has not been determined. Although there is 
clear-cut evidence for the presence of circumstellar gas and 
dust (see e.g. Waters \& Waelkens 1998), there is no consensus on 
the {\it geometry} of this material (see e.g. Pezzuto, Strafella \& 
Lorenzetti 1997 for a summary of conflicting views). Ideally one would 
like to solve this issue by directly imaging the environments around these 
objects. Although there are indications that some 
intermediate mass stars present geometries deviating from 
spherical symmetry (e.g. Mannings \& Sargent 1997; Grady et al. 1999; Shepherd,  
Claussen \& Kurtz 2001), the conclusions reached are sometimes contradictory.
For instance in the case of AB~Aur, the brightest and best-studied 
Herbig Ae/Be star, Mannings \& Sargent infer an almost edge-on 
geometry, i~=~76\degree, while Grady et al. claim i~$\leq$~45\degree. 
Furthermore, most of these imaging studies have been performed in the 
millimetre and radio regimes, probing the geometry on a relatively large 
scale, between about 100 -- 1000 AU. To make progress in the area of 
star formation for the higher mass stars, observations exploring the 
innermost regions around these objects are badly needed. Until 
sub-milli-arcsecond imaging becomes a reality, linear spectropolarimetry 
across emission lines such as H$\alpha$ is the most powerful technique 
for doing so, as the scattering of starlight on electrons in an ionised 
medium occurs within a few stellar radii (Rudy 1978; Cassinelli, Nordsieck, \& Murison, 1987).

This application of spectropolarimetry was first established in studies of 
classical Be stars (Clarke \& McLean 1974; Poeckert 1975).  In its 
simplest form, it is based on the expectation that H$\alpha$ photons arise 
over a much larger volume than the stellar continuum photons.
For this reason the line photons undergo fewer scatterings off e.g. a
circumstellar disc than the continuum photons do. Consequently 
the emission line flux will be much less polarised than the continuum.
In this situation, a change 
in polarisation across the line profile is expected -- we 
refer to this {\it de}polarisation as the classical line effect (because
of its first appearance in observations of classical Be stars).
The high incidence of these line depolarisations among members of this 
object class (26 out of 44 in Poeckert \& Marlborough 1976) indicated 
that the envelopes of classical Be stars are not spherically symmetric.  
These findings are now taken as compelling evidence that classical 
Be stars are embedded in circumstellar discs (see e.g. Waters \& Marlborough 1992).  

Since then, H$\alpha$ spectropolarimetry has been performed on a variety 
of other strong line emitting objects, such as B[e] stars 
(Zickgraf \& Schulte-Ladbeck 1989; Magalh$\tilde{\rm a}$es et al. 1992), 
Luminous Blue Variables (e.g. Schulte-Ladbeck et al. 1994), 
novae (Bjorkman et al. 1994) and supernovae (e.g. Cropper et al. 1988). 
In the case of pre-main sequence stars the technique has recently been 
applied to a sample of 9 Herbig Be stars by Oudmaijer \& Drew (1999; 
hereafter OD99). The main outcome of this pilot study was 
that a line effect was detected in about half the Herbig Be stars.  
It is possible that the OD99 detection rate is consistent with discs 
around all these stars being viewed at angles ranging from face-on to 
edge-on.  But to be able to make a proper distinction between viewing 
angle and intrinsic geometry effects, one needs to obtain a larger sample.  
We pursue this goal by increasing the sample of spectropolarimetric data 
to about 25 Herbig stars in total. 

In addition to increasing the sample, we extend our data set toward the 
later spectral type Herbig Ae (HAe) stars, where we may be able to detect 
spectropolarimetry characteristics differing from the earlier type Herbig Be 
(HBe) stars. A switch in phenomenology may be expected to occur at some point 
working down the stellar mass range, as different physical mechanisms are 
likely to play a role at different spectral types.  For instance, radiation 
pressure forces are likely to play a role for the higher luminosity stars at 
the early B types, whereas magnetic fields may become more dynamically 
prominent at the later A types.  The magnetically-channelled accretion model 
(Ghosh \& Lamb 1979) that is commonly applied to the lowest mass 
pre-main-sequence T~Tauri stars (by e.g. Bouvier, Forestini, \& Allain, 1997) may also be a 
suitable model as early as spectral type A (Pontefract et al. 2000).  If it 
does operate, the inner accretion disc around the star is truncated by the 
magnetic field, and the depolarisation effect may then be absent 
because the inner hole will necessarily lead to reduced intrinsic continuum 
polarisation.  Alternatively, the channelled accretion may produce a 
relatively bright and compact source of H$\alpha$ emission that may be 
scattered within the accretion column itself or within the disrupted 
disc.  This in turn may yield a polarisation signature at H$\alpha$ 
that is more complex than the simple depolarisation effect (see 
McLean 1979; Wood, Brown \& Fox 1993).  The second goal of this paper 
is accordingly to explore how the H$\alpha$ spectropolarimetric signature behaves 
as a function of Herbig star spectral type.  

The paper is organised as follows. In Sect.~\ref{s_obs} we discuss the way
in which our sample was selected, how the observations were performed, and 
reduced. In Sect.~\ref{s_results} we first present the continuum polarisation 
results for the whole sample and subsequently 
focus on the spectropolarimetry. We outline spectropolarimetric behaviour 
that can be expected, and develop tools to characterise the observations. 
We then discuss the individual targets, where we proceed by separating the 
discussion on the Herbig Be stars (Sect.~\ref{s_hbe}) from the Herbig Ae 
stars (Sect.~\ref{s_hae}). The two groups are subsequently compared in 
Sect.~\ref{s_diff}. In Sect.~\ref{s_disc} we summarise the main outcomes of 
our study and discuss the possible interpretations. 

\begin{table*}
\begin{minipage}{\linewidth}
\renewcommand{\thefootnote}{\thempfootnote}
\caption{Herbig Ae/Be targets. {\it V} magnitudes (column 3) and Spectral types (column 4) 
are taken from {\sc simbad}. The integration times denote the total exposures (column 6).
The errors in our polarisation data (column 7) are of the order of 0.01\% based 
on photon-statistics only. Yet, systematic (external) errors in the 
polarisation are estimated to be 0.1\%. The errors in the Position Angle $\theta$ (column 8) 
are less than a degree. Column (9) gives estimates of the sky PA derived 
from line excursions; note that the presumed discs are supposed to lie perpendicular 
(=90\degree) to this determined sky PA.}
\label{t_cont}
\begin{tabular}{llrlclcrc}
\hline
Name          & HD Number  & {\it V} &Spec. Tp  & Date & Exposure(s)  & $P_{\rm cont}^{\rm R}$ (\%) & $\Theta_{\rm cont}^{\rm R}$ (\degree) & $\Theta_{\rm intr}^{\rm R}$ (\degree)\\
\hline
MWC 137       &              & 11.2     & Be     &20-12-99         & 24x100  & 6.07 & 160  & 115 $\pm$ 10\\
MWC 1080      &              & 11.6     & B0     &18-12-99             & 8x500   & 1.73 & 77   & \\
MWC 166       & HD 53367     &  7.0     & B0     &18-12-99             & 16x150  & 0.20 & 34   &  \\
BD+40 4124    &              & 10.7     & B2     &19-12-99         & 24x100  & 1.21 &  8   & 173 $\pm$ 8\\
MWC 361       & HD 200775    &  7.4     & B2Ve   &18-12-99             & 16x100  & 0.81 & 96   &  \\
IL Cep        & HD 216629    &  9.3     & B2IV-Vne&19-12-99            & 16x100  & 4.24 & 102  &   \\
MWC 147       & HD 259431    &  8.8     & B6pe   &18-12-99             & 24x50   & 1.05 & 100  &  \\
MWC 158$^1$   & HD 50138     &  6.6     & B9     &18-12-99         & 16x20   & 0.65 & 59   & 45 $\pm$ 5\\
              & HD 58647     &  6.8     & B9IV   &19-12-99             & 16x75   & 0.14 & 127  & \\
AS 477        &              & 10.2     & B9.5Ve &19-12-99             & 16x150  & 0.43 & 56   &  \\
MWC 120       & HD 37806     &  7.9     & A0     &19-12-99             & 16x50   & 0.35 & 76   & \\
KMS 27        & HD 37357     &  8.9     & A0     &19-12-99             & 16x200  & 0.13 & 52   & \\
MWC 789       & HD 250550    &  9.6     & A0     &20-12-99         & 32x100  & 0.92 & 174  & 178 $\pm$ 10\\
AB Aur        & HD 31293     &  7.1     & A0pe   &18-12-99         & 16x150  & 0.11 & 54  & 160 $\pm$ 5\\
SV Cep        &              & 10.1     & A0     &20-12-99             & 8x500   & 0.61 & 67  &  \\
XY Per        & HD 275877    &  9.4     & A2II   &20-12-99             & 8x300   & 1.60 & 132 &   \\
MWC 480       & HD 31648     &  7.7     & A2     &19-12-99             & 16x75   & 0.38 & 52  &  \\
              & HD 244604    &  9.4     & A3     &19-12-99             & 16x200  & 0.44 & 119 &   \\
MWC 758       & HD 36112     &  8.3     & A3     &19-12-99             & 16x100  & 0.07 & 179 &   \\
T Ori         &              &  9.5     & A3V    &20-12-99             & 8x500   & 0.39 & 97  &  \\
              & HD 245185    & 10.0     & A5     &20-12-99             & 8x500   & 0.21 & 168 &   \\
              & HD 35929     &  8.1     & A5     &19-12-99             & 16x200  & 0.12 & 51  &  \\
CQ Tau$^2$        & HD 36910     & 10.7     & F2IVe  &20-12-99             & 8x500   & 0.27 & 83 &   \\
\hline
\end{tabular}
\\
\noindent
$^1$ MWC~158 does not appear in Table 1 of Th\'e et al. (1994). The data have been 
included for completeness.\\
\noindent
$^2$ Note that the spectral type for CQ~Tau as given by Th\'e et al. (1994) 
is between late A and early F (A8 -- F2), and\\
\hspace*{0.2cm} 
is indeed considered to be a member of the Herbig Ae/Be group.\\
\end{minipage}
\end{table*}

\section{Observations}
\label{s_obs}

\subsection{Sample selection}

Our target stars were selected from the Herbig Ae/Be 
catalogue of Th\'e, de Winter \& Perez (1994). Note that in the 
following only objects appearing in Table~1 from their catalogue
will be discussed and considered in the statistics as ``genuine'' Herbig 
stars. Herbig (1960) originally selected these stars on the basis 
of three criteria, but more recently Waters \& Waelkens (1998) stressed 
the importance of a dusty infrared excess. 
Most of the objects in Table~1 by Th\'e et al. follow the criteria, but a few 
objects may not have a dusty infrared excess. 
Yet, it is the largest representation of a genuine intermediate mass young 
population that is currently available.     

All the targets from this list that we were then able to observe are 
listed in our Table~\ref{t_cont}, alongside their spectral types and
V magnitudes, as listed in {\sc simbad}.  It is important to note 
that the targets were not selected with any foreknowledge of 
flattened circumstellar geometries, but were selected because of their 
relative brightness ($V \la 11$) and their position on the sky.

\subsection{Spectropolarimetric observations}

The linear spectropolarimetric data were obtained during the 
nights of 1999 December 18 -- 20 using the ISIS spectrograph 
mounted on the Cassegrain focus of the 4.2-metre William 
Herschel Telescope (WHT), La Palma. For all observations a slit width
of $1.2 \arcsec$ was used. 
Although the sky was relatively clear, the seeing was rather poor 
($\sim 1.5 - 3 \arcsec$).
We used a 1024 $\times$ 1024 pixel TEK-CCD detector
with the 1200R grating, which 
yielded a spectral coverage from 6370 -- 6760 \ang.
This coverage corresponds to a spectral resolution
of approximately 35 \kms~around \ha\ (as measured from arc line fits).

To be able to analyse the linearly polarised component in the spectra, 
ISIS was equipped with the appropriate polarisation optics.
This consisted of a rotating half-wave plate and a calcite block to rotate 
and separate the light into two perpendicularly polarised light waves -- 
respectively the ordinary (O) and extraordinary (E) rays.  
Two holes in the Dekker also allow for simultaneous observations 
of the object and the sky.  Hence, for each exposure, four spectra are 
recorded: the O and E rays of both the target and sky.  
One complete observation set consists of a series of four 
exposures at half-wave plate position angles of, in chronological
order, 0\degree, 45\degree, 22.5\degree, and 67.5\degree~to 
obtain the linear Stokes parameters $Q$ and $U$. 
After each of these sets of four frames were obtained, the holes
in the Dekker for the target and the sky were interchanged 
to compensate for any kind of asymmetry in either the detector 
or the instrument.   To prevent the CCD from saturating
on the peak of \ha, shorter exposure times were adopted for those 
objects with strong \ha\ emission.
Finally, polarised and zero-polarisation standards were observed 
regularly, revealing an intrinsic instrumental 
polarisation of the order of 0.1 per cent, which we do not
correct for, as our aim is only to investigate \ha\ polarimetric 
signatures relative to the continuum.

\subsection{Data reduction and calibrations}

The reduction of the data was carried out using {\sc iraf} (Tody 1993).
The data reductions steps included the usual (i) bias-subtraction, 
(ii) flat-fielding, (iii) cosmic ray removal, (iv) extraction of the spectra 
and (v) wavelength calibrations of the O and E spectra. 

The E and O ray data were then imported into the polarimetry package 
{\sc ccd2pol}, incorporated in the {\sc figaro} software package maintained 
by {\sc starlink}. The Stokes parameters $Q$ and $U$ were determined, leading
to the percentage linear polarisation $P$ and its Position Angle (PA) $\theta$ in the 
following way:

\begin{equation}
P~=~\sqrt{(Q^2 + U^2)}
\end{equation}
\begin{equation}
\theta~=~\frac{1}{2}~\arctan(\frac{U}{Q})
\end{equation}
Note that a PA of 0\degree, i.e. North-South, on the sky is represented by
a vector that lies parallel to the positive $Q$ axis, whereas $\theta$\,=\,90\degree\ (i.e. 
East-West) is positioned in the negative $Q$ direction. Positive and negative $U$ axes 
thus correspond to position angles of respectively 45\degree and 135\degree.

The data were subsequently analysed using {\sc polmap}. 
This package was also used to remove an instrumental, wavelength-dependent 
ripple (see Harries \& Howarth 1996 for a discussion on the derippling procedure). 
The success of this last step in the data extraction is occasionally less 
than complete, such that we will have to comment on a few instances 
where some ripple seems to have remained. 
The achieved (relative) accuracy of the polarisation data is in principle only 
limited by photon-statistics and can be very small (typically 0.01 \%). 
However, the quality and the amount 
of data taken on spectropolarimetric standard stars is at present not 
yet sufficient to reach absolute accuracies below 0.1\% (see the manual by 
Tinbergen \& Rutten, 1997). 
In most parts of the paper the spectropolarimetric data are shown binned
to a constant error of 0.05\% polarisation and therefore the presented 
polarisation spectra exhibit a spectral resolution that depends on the total 
number of counts (i.e. along the \ha\ profile, the highest resolution is 
achieved at the wavelength of the emission peak). In a few cases, we choose
to slightly modify the error per bin in order to achieve 
the best compromise between minimising the error per bin and 
resolving the line profiles.

Just as we do not correct for instrumental polarisation,
no correction for interstellar polarisation (ISP) is made either.
This is because the ISP only adds a wavelength-independent 
polarisation vector to all observed points when plotted in 
$(Q,U)$ space. Precisely because of this wavelength independence, we 
will particularly rely on the $(Q,U)$ plane representation of the data 
in classifying the type of \ha\ polarisation signature.

\begin{table*}
\begin{minipage}{\linewidth}
\renewcommand{\thefootnote}{\thempfootnote}
\caption{The \ha\ line results. 
The errors on the equivalent widths of the \ha\ lines (column 2) 
are below 5\%, the errors on $\Delta \lambda(\rm Pol)$ (column 5) 
and $\Delta \lambda(I)$ are determined at Full Width Zero Intensity (FWZI) 
and are about 10\%.
$\Delta \lambda(\rm Pol)$ has been defined as the width over 
which the polarisation changes. In the case where the widths in PA 
and \%Pol are unequal, we take the largest of the two.  
The fractional width \ratpol\ is given in Column (6).  
The followed recipe with regard to the depolarisation question (column 7) 
is described in the text. Column (8) represents the morphology in $(Q,U)$ 
space. 
\label{t_line} }
\begin{tabular}{lcccccllc}
\hline
Object & \ha\ EW(\ang) & Line/Cont &  Line  & $\Delta \lambda(\rm Pol)$ & ($\frac{\Delta \lambda(\rm Pol)}{\Delta \lambda(I)}$) & Depolarisation? & $(Q,U)$ & Spectral\\
       &  & Contrast  & Effect? & (\ang) & & & behaviour  & Type\\
\hline
MWC 137    & $-$404 & 70    & yes &55 &0.92   & yes & exc     &B\\   
MWC 1080   & $-$101 & 18    & yes &50 &1.11   & yes & smear       &B0\\      
MWC 166    & $-$2.8 & 1.6   & no? & -- & --   & --    & --    &B0\\
BD+40 4124 & $-$113 & 17    & yes &30 &0.75   & yes & exc     &B2\\
MWC 361    & $-$63  & 8.5   & yes &45 &1.13   & yes(+) & exc     &B2\\   
IL Cep     & $-$20  & 3.5   & no  &-- &--     & --  & --     &B2\\   
MWC 147    & $-$60  & 13    & yes &18 &0.50   & no  & exc     &B6\\   
HD 58647   & $-$8.6 & 2.4   & yes &10 &0.67   &  no  & loop    &B9\\   
MWC 120    & $-$29  & 6.3   & yes &13 &0.65   & no     & loop    &A0\\
MWC 789    & $-$46  & 10    & yes &21 &0.95   & no  & loop+exc&A0\\
AB Aur$^1$ & $-$40  & 8.0   & yes &17 &1.00   & no  & loop+exc&A0\\     
XY Per     & $-$6.7 & 2.3   & yes &15 &0.88   & no  & loop    &A2\\   
MWC 480    & $-$21  & 4.8   & yes &12 &0.67   & no  & loop    &A2\\   
HD 244604  & $-$14  & 4.5   & yes &10 &0.56   & no  & loop?   &A3\\ 
MWC 758    & $-$17  & 4.0   & no  &-- &--     & --  & --      &A3\\   
T Ori      & $-$12  & 3.9   & yes &5  &0.26   & no  & loop    &A3\\   
HD 245185  & $-$17  & 3.7   & yes &10 &0.56   & no  & loop    &A5\\   
HD 35929   & $-$3.2  & 2.0  & no  &-- &--     & --  & --      &A5\\   
CQ Tau     & $-$2.7  & 1.6  & yes &3  &0.17   & no  & loop    &F2\\   
  & & & & & & & & \\
\hline 
   & & & & & & & & \\
MWC 158    & $-$54  & 12    & yes &25 &1.0    & no    & exc     &B9\\
AS 477$^2$ & $-$19  & 4.4   & no? &-- &--     & --  & --      &B9.5\\  
KMS 27$^3$ & $-$7.2 & 2.5   & yes &-- &--     & --  & --      &A0\\   
SV Cep$^2$ & $-$13  & 2.5   & yes &-- &--     & --  & --      &A0\\   
   & & & & & & & & \\
\hline
\end{tabular}
\\
$^1$ For AB~Aur an improved data reduction sequence was followed in comparison 
to the data presented in Pontefract\\
\hspace*{0.2cm}
et al. (2000); the conclusions reached in Pontefract et al. remain valid.\\
$^2$ The data on AS~477 and SV~Cep suffer from too low photon counts 
to interpret them.\\   
$^3$ The smear in KMS~27 is due to the broad photospheric absorption wings. 
\end{minipage}
\end{table*}

\section{General Results}
\label{s_results}

We begin with a brief presentation of the continuum 
polarisation results (Sect.~\ref{s_cont}), before we focus on 
the spectropolarimetry (Sect.~\ref{s_spec}). The discussion of  
individual targets is performed in the next section (Sect.~\ref{s_ind}).

\subsection{Continuum Polarimetry}
\label{s_cont}
 
The linear polarisation of Herbig Ae/Be star continua can be attributed to 
the scattering of stellar photons off matter within
an asymmetric circumstellar geometry. In addition there may be 
a substantial interstellar contribution.  It is not yet settled which 
polarigenic agent causes the circumstellar polarisation, but the
most likely candidates are dust particles and electrons. 
Whilst the resolution of this question is both interesting and
significant, we leave it for the future, given our present focus on 
the implications of the line polarisation morphology for the circumstellar 
geometry.

The measured continuum polarisations for both the HBes and HAes are 
summarised in Table~\ref{t_cont}. The mean percentage polarisation 
and position angle obtained for each object appear in columns (7) and (8). 
In addition, for cases where an excursion in $(Q,U)$ space (see later) is present, 
we estimate the sky PA from these line excursions, and 
present them in column (9) for completeness. (Note that the presumed 
discs are supposed to lie perpendicular to these sky PA estimates).

Polarisation variability is a well-known and common phenomenon (Vrba, 
Schmidt \& Hintzen, 1979; Grinin 1994). In general terms, it is viewed as 
caused by time variable amounts and/or projected distributions of 
circumstellar gas and dust.  Accordingly we do not look for perfect 
agreement between our continuum polarisation measurements and those in 
earlier literature.  Nevertheless, the values of the polarisation 
quantities, \%Pol and PA, are generally consistent with previous 
continuum measurements in the R band (Petrova \& Shevchenko 1987; 
Jain \& Bhatt 1995; OD99; and Oudmaijer et al. 2001). 

\subsection{Spectropolarimetry}
\label{s_spec}

The observed \ha\ characteristics of each Herbig star are listed 
in Table~\ref{t_line}. The first few columns list the \ha\ parameters 
deducible from the ``intensity'' (Stokes I) profile only, namely the equivalent 
width (EW; column 2) and the ratio of the peak line counts over the 
continuum counts (column 3). These parameters can be used 
as indicators for the amount of ionised gas in the environments of these 
objects. As one would expect, these quantities show a 
general decline toward later spectral types. 

The spectroscopic line shapes of the Herbig stars vary 
from pure emission, double-peaked emission, P Cygni type, through 
to inverse P Cygni type. 
We will see later that these spectroscopic line profiles do not 
necessarily correlate with the spectropolarimetry. In fact, spectropolarimetric 
behaviour seems to be rather more homogeneous than Stokes I spectroscopy. 
This is illustrative of how the polarised light contribution to the total
light has a more selective scattering origin. It is an independent dimension
of information on the source geometry.
In other words, spectropolarimetric line profiles offer the opportunity 
to study physical phenomena that are unobservable in normal spectroscopy.

Columns (4) to (8) in the 
Table concern spectropolarimetric properties.
Plots of the spectropolarimetric data for the Herbig Ae/Be stars 
are presented in the different panels of Figs.~\ref{f_hbe} 
and \ref{f_hae}.
For all objects the polarisation spectra are
presented as triplots (consisting of Stokes I, $P$, and 
$\theta$), as well as loci in $(Q,U)$ space.

\subsubsection{Spectropolarimetric expectations}

\begin{figure}
\begin{center} 
\mbox{\epsfxsize=0.38\textwidth\epsfbox{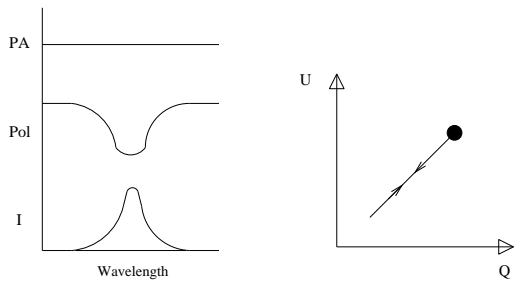}}
\caption{Cartoon: depolarisation across \ha; seen in a 
triplot and a $(Q,U)$ diagram. A typical Stokes I emission 
is shown in the lower panel of the triplot, 
the \%Pol in the middle panel, while the Position Angle is 
sketched in the upper panel of the triplot. Note that the 
depolarisation across \ha\ is as broad as the Stokes I emission. Depolarisation
translates into $(Q,U)$ space as a linear excursion.}
\label{f_depolariz}
\end{center}
\end{figure}

Consider a star embedded in an at least partly ionised circumstellar 
medium. If (part of) the \ha\ photons are formed over a large volume, 
and the projected distribution of Thomson-scattering free electrons 
is sufficiently flattened on the sky, the line emission itself is
unpolarised while the continuum shows some linear polarisation. 
One may then expect to detect a 
depolarisation line effect (as found in classical Be stars).  Where this 
happens one would expect to see a change in \%Pol across the line 
that more or less ``follows'' the shape of the Stokes I intensity profile.
Hence the polarisation change is expected to be as broad as the
line emission itself (see the middle panel of the triplot sketch 
in Fig.~\ref{f_depolariz}).
In this case the depolarisation across \ha\ is only seen in the \%Pol, 
but in practise (due to the vector addition of the foreground polarisation), 
it may be observable in the PA as well or instead. 
In any case, when mapped onto $(Q,U)$ space, the line 
wavelength points will emerge from the continuum knot 
as a linear excursion (as seen in the $(Q,U)$ plane sketch of 
Fig.~\ref{f_depolariz}). 
 
\begin{figure}
\begin{center} 
\mbox{\epsfxsize=0.38\textwidth\epsfbox{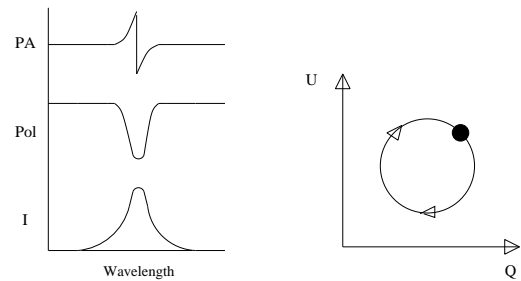}}
\caption{Cartoon: compact \ha\ scattered in rotating disc. Note 
that the polarisation signatures are relatively narrow compared 
to the Stokes I emission. The flip in PA is associated with a loop
in $(Q,U)$ space.}
\label{f_intr}
\end{center}
\end{figure}

\begin{figure}
\begin{center} 
\mbox{\epsfxsize=0.38\textwidth\epsfbox{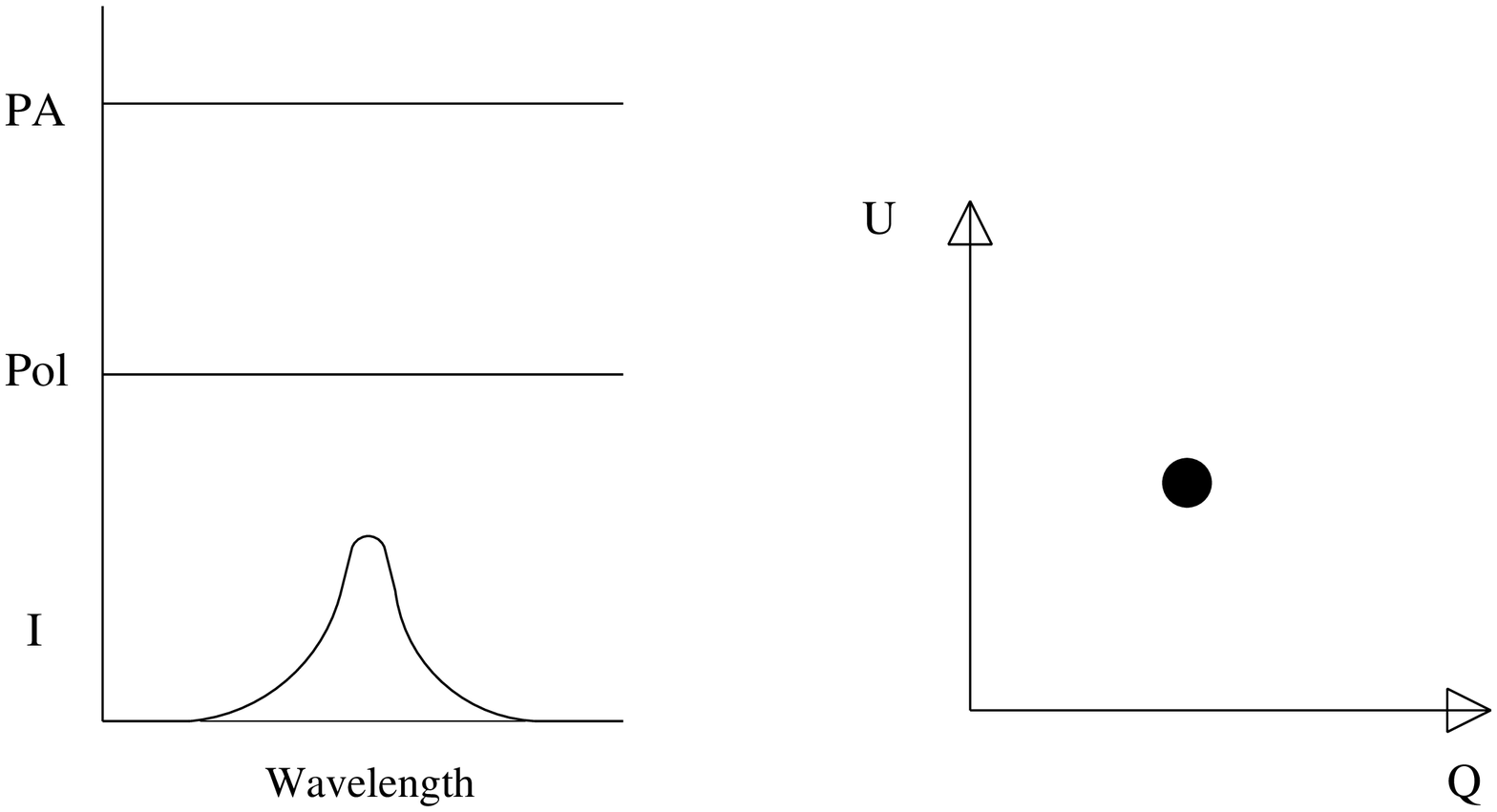}}
\caption{Cartoon: ``no line effect''}
\label{f_noline}
\end{center}
\end{figure}

If, on the other hand, (part of) the \ha\ line photons originate from a
{\it compact} rather than an extended source, one may pick out 
more subtle line effects, as have for instance been modelled by Wood et al. 
(1993; see also McLean 1979).  In the presence of a compact \ha\ source one 
can envision a variety of line effects that will map into $(Q,U)$ space in as
many ways.  The patterns seen will depend on the specific geometry and 
bulk motions of the scattering particles surrounding the compact \ha\ 
source.  If the scattering occurs predominantly in a rotating disc-like 
configuration, the subsequent breaking of left-right reflection symmetry
in the velocity field leads to a changing PA with wavelength (McLean 
1979), which appears as a ``loop'' in $(Q,U)$ space (see the 
cartoon in Fig.~\ref{f_intr}). If the scattering 
however occurs in an expanding (as opposed to a rotating) medium, 
there is no breaking of the left-right symmetry, and the track 
in $(Q,U)$ space will appear to be more ``linear''. 
In case of a rotating {\it and} an expanding medium, 
the tracks in $(Q,U)$ space are expected to appear as a combination
of the two and thus be more or less ``hook-shaped'' (see 
Schulte-Ladbeck et al. 1994; Oudmaijer et al. 1998). In all of these 
compact \ha\ cases there is the possibility that the width of the polarisation
profile across \ha\ does not coincide with the width of the Stokes I 
intensity profile (see the triplot sketch in Fig.~\ref{f_intr}).

Finally, where the geometry on the sky is circular, as is the 
case for either spherically symmetric geometry or a face-on disc, there is 
no change in PA and/or \%Pol across the line.  In this circumstance
and in the ideal limits of noiseless data and no wavelength dependence of
foreground polarisation, all points in $(Q,U)$ space would collapse 
into a single ``dot'' (see Fig.~\ref{f_noline}).  
For some objects described below there is no \ha\ line effect apparent 
in the triplot, but yet due to polarisation changes in the continuum  
or broad photospheric absorption wings the ``dot'' smears to a short 
fuzzy line, called a ``smear''.

\begin{figure*}
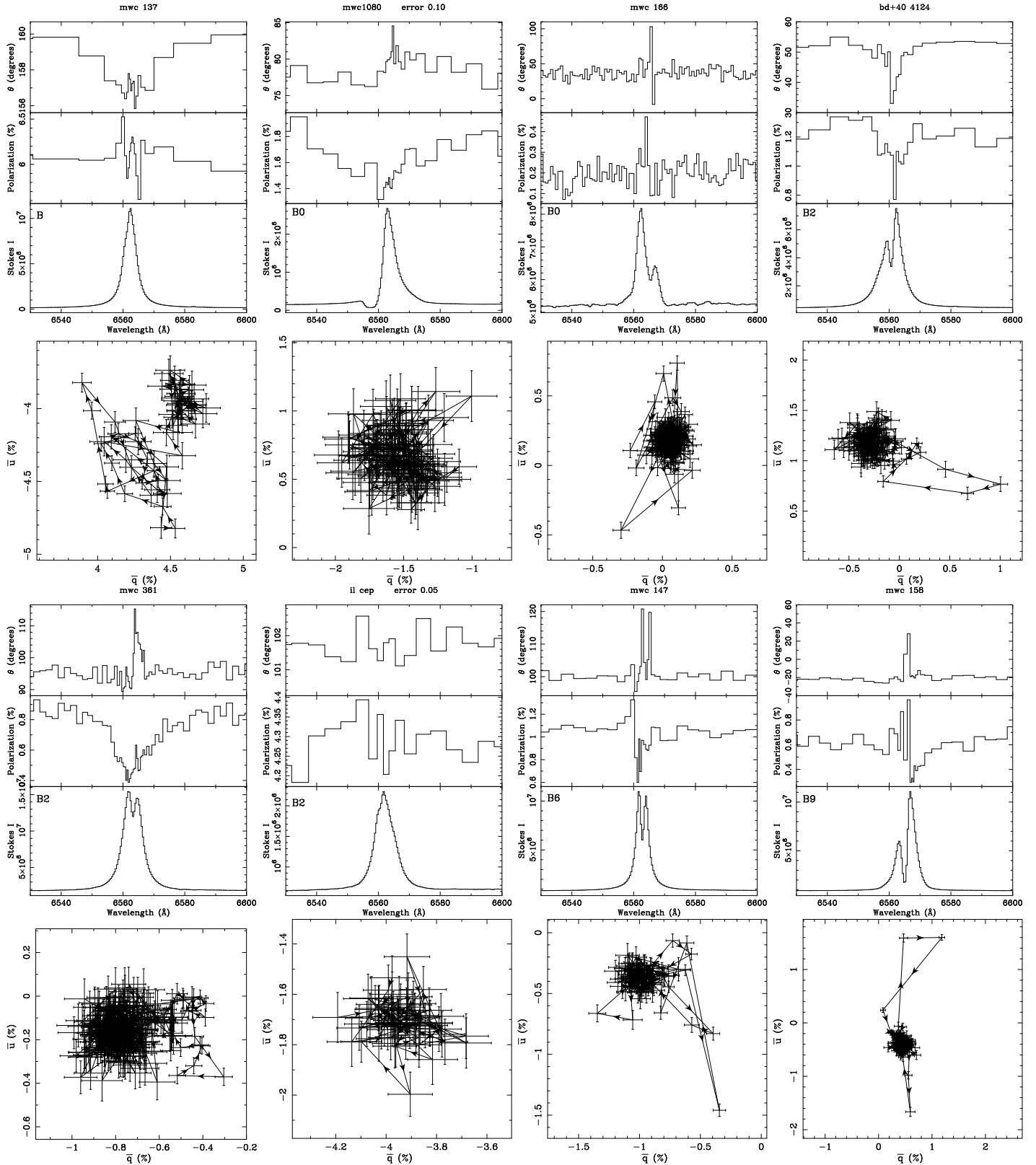

\mbox{
\epsfxsize=0.26\textwidth\epsfbox{mwc137.ps}
\epsfxsize=0.26\textwidth\epsfbox{mwc1080.ps}
\epsfxsize=0.26\textwidth\epsfbox{mwc166.ps}
\epsfxsize=0.26\textwidth\epsfbox{bd+40.ps}
}
\mbox{
\epsfxsize=0.26\textwidth\epsfbox{qu_mwc137.ps}
\epsfxsize=0.26\textwidth\epsfbox{qu_mwc1080.ps}
\epsfxsize=0.26\textwidth\epsfbox{qu_mwc166.ps}
\epsfxsize=0.26\textwidth\epsfbox{qu_bd+40.ps}
}
\mbox{
\epsfxsize=0.26\textwidth\epsfbox{mwc361.ps}
\epsfxsize=0.26\textwidth\epsfbox{ilcep.ps}
\epsfxsize=0.26\textwidth\epsfbox{mwc147.ps}
\epsfxsize=0.26\textwidth\epsfbox{mwc158.ps}
}
\mbox{
\epsfxsize=0.26\textwidth\epsfbox{qu_mwc361.ps}
\epsfxsize=0.26\textwidth\epsfbox{qu_ilcep.ps}
\epsfxsize=0.26\textwidth\epsfbox{qu_mwc147.ps}
\epsfxsize=0.26\textwidth\epsfbox{qu_mwc158.ps}
}
\caption{
Polarisation spectra of Herbig Be stars and their 
$(Q,U)$ diagrams (below the triplots). 
For each star the Stokes I spectrum is shown 
in the lowest panel of the triplot, 
the \%Pol in the middle panel, 
while the PA ($\theta$; see Eq.~2) is 
plotted in the upper panel. The data are 
rebinned such that the 1$\sigma$ error in the polarisation
corresponds to 0.05\% as calculated from photon statistics. In 
case of exceptions on this, the polarisation errors are given 
on top of the triplots. The spectral types 
are indicated in the Stokes I plots for convenience.
The lower plot represents the normalised Stokes parameters
$u = U/I$ against $q = Q/I$.
Note that the binning errors in these $(Q,U)$ plots are 
consequently multiplied by a factor of two (compared to the binning 
errors in the triplots) to present a well-defined continuum.
\label{f_hbe}
}
\end{figure*}

\addtocounter{figure}{-1}%

\begin{figure*}
\mbox{
\epsfxsize=0.26\textwidth\epsfbox{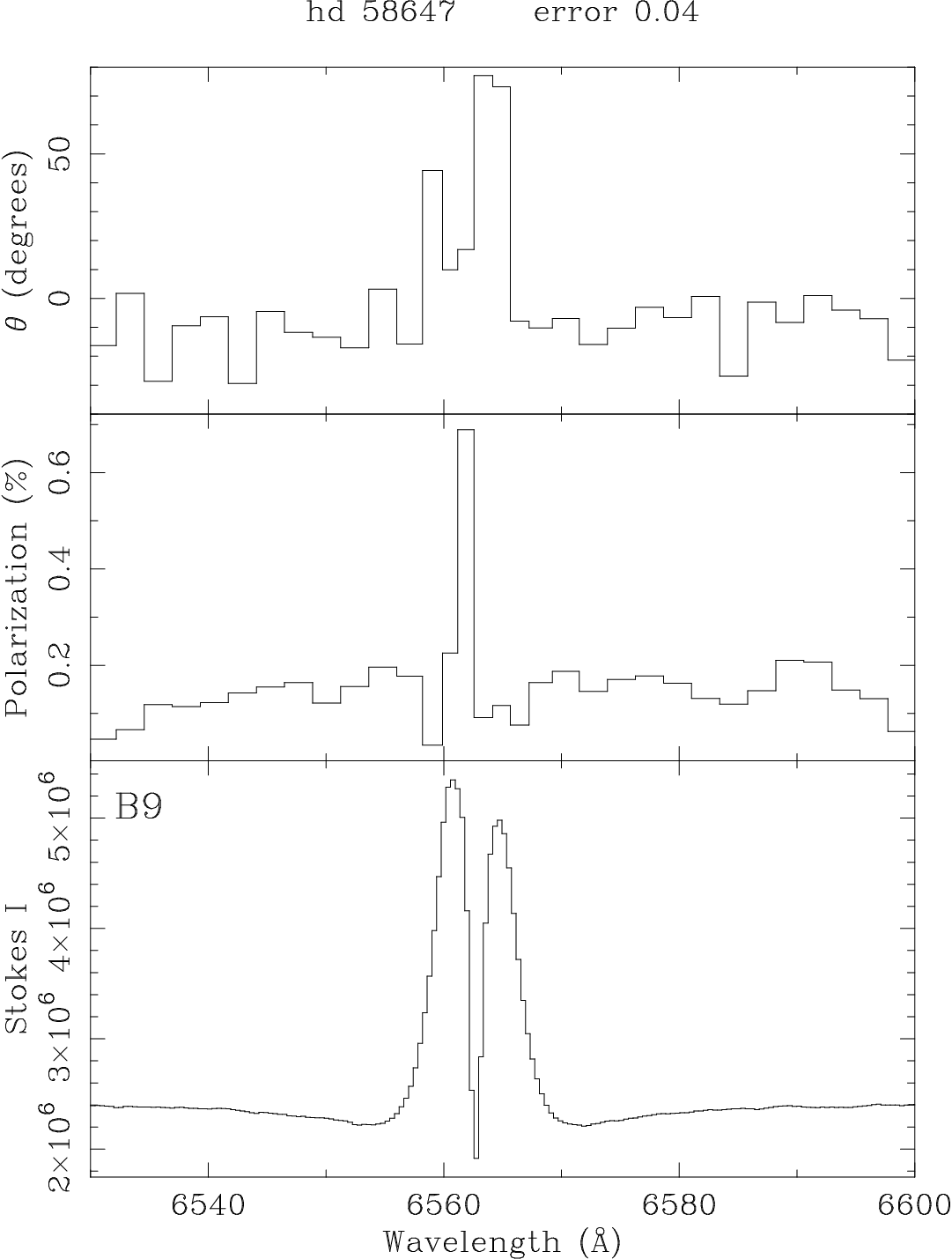}
\epsfxsize=0.26\textwidth\epsfbox{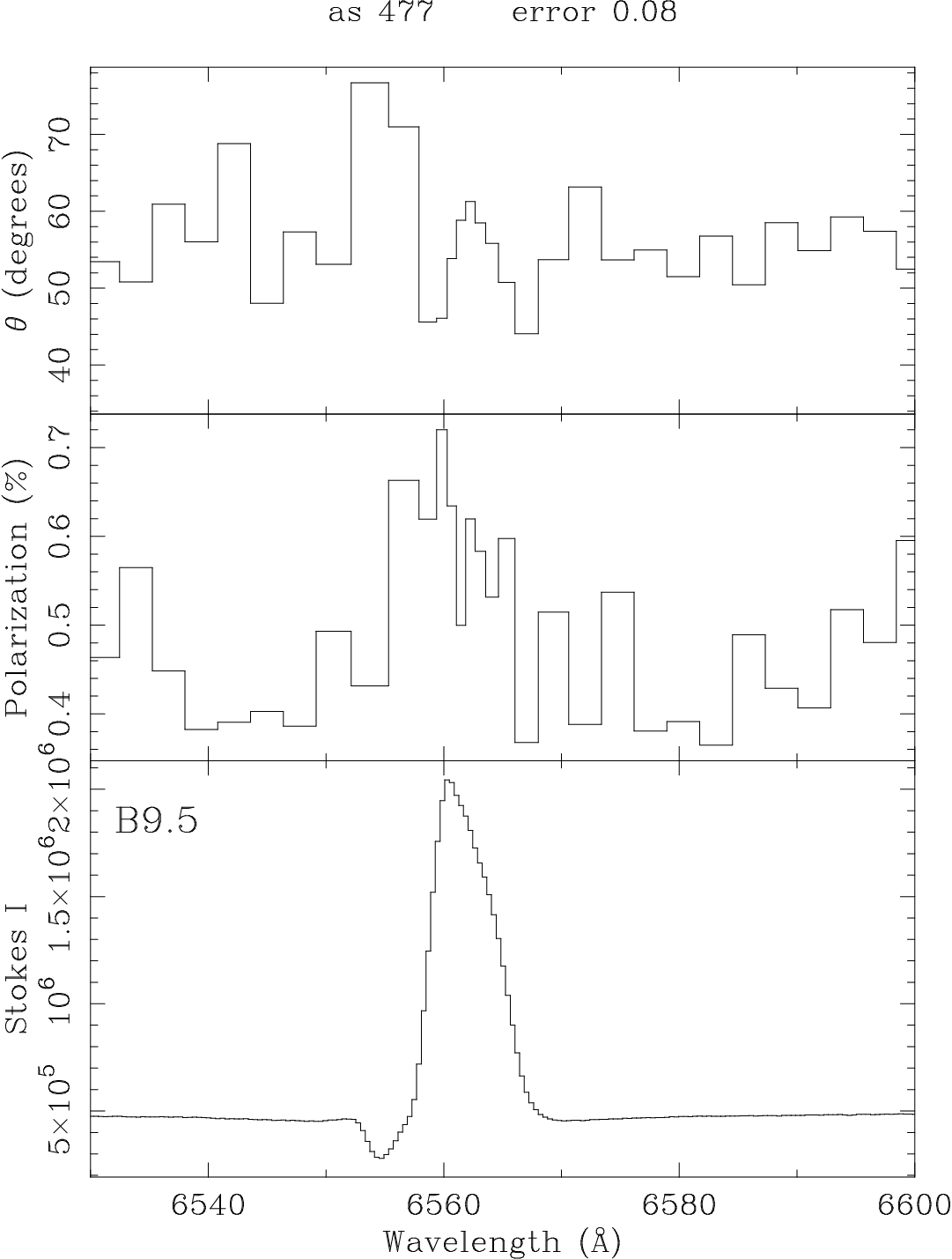}
}
\mbox{
\epsfxsize=0.26\textwidth\epsfbox{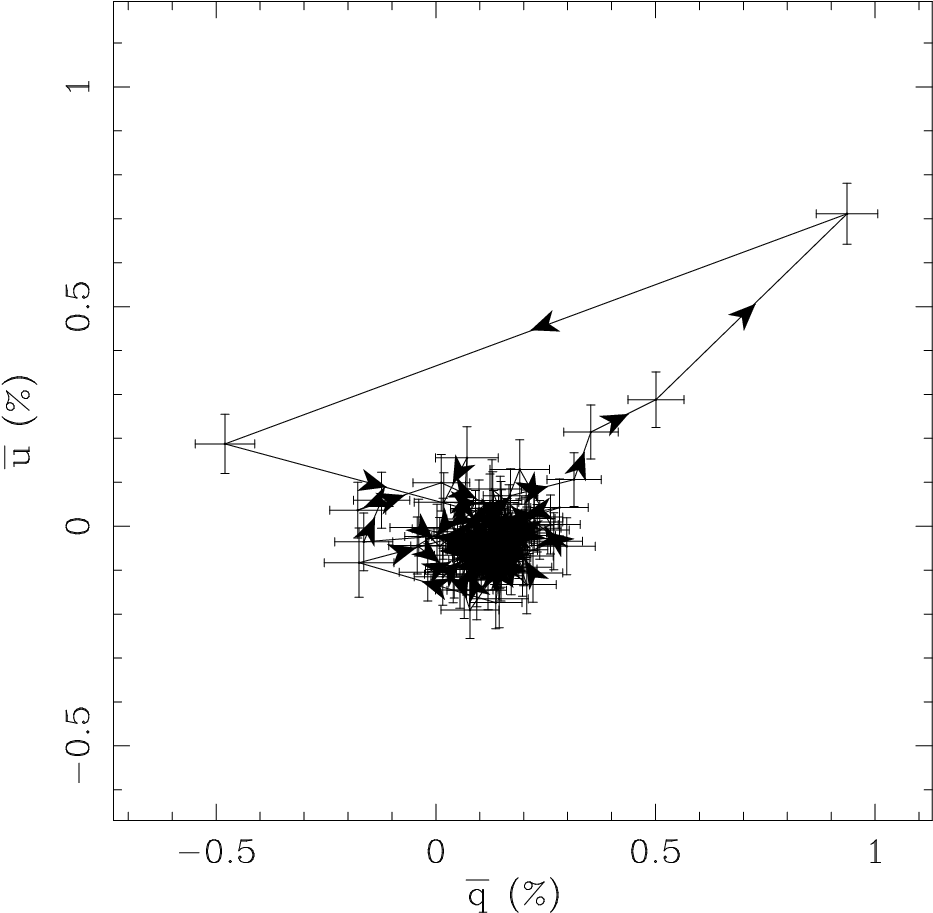}
\epsfxsize=0.26\textwidth\epsfbox{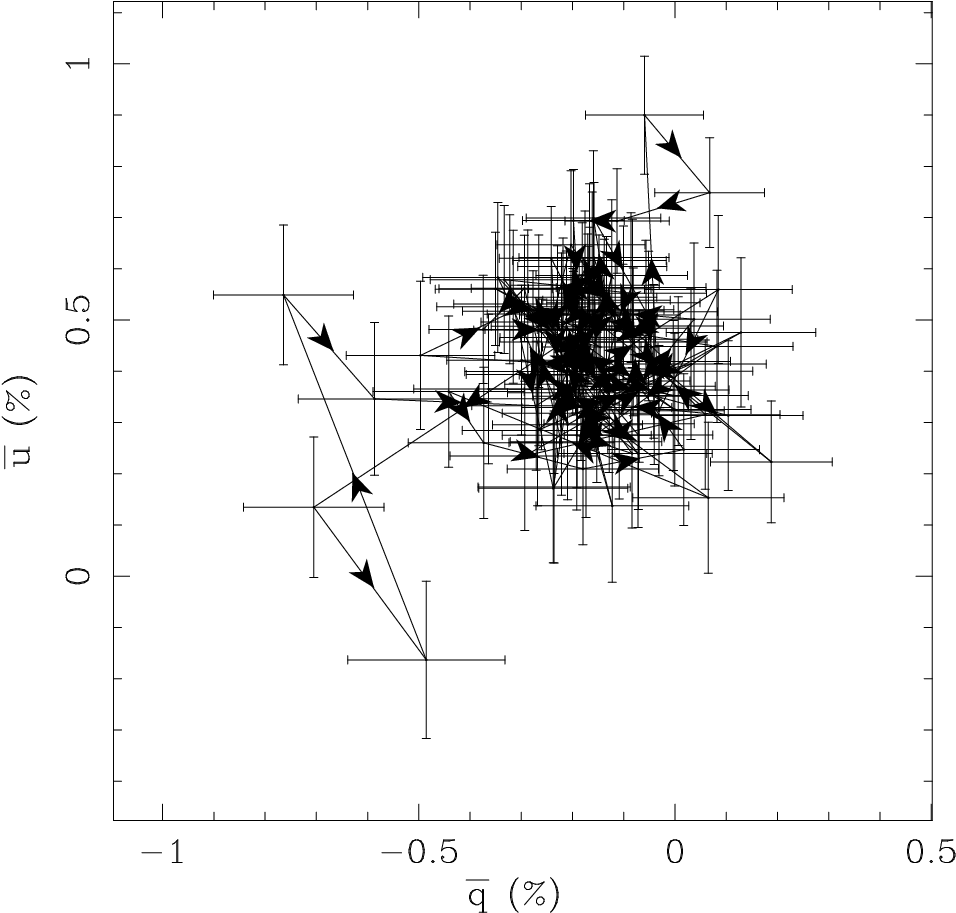}
}
\caption{Continued}
\label{f_hbe2}
\end{figure*}

\begin{figure*}
\mbox{
\epsfxsize=0.26\textwidth\epsfbox{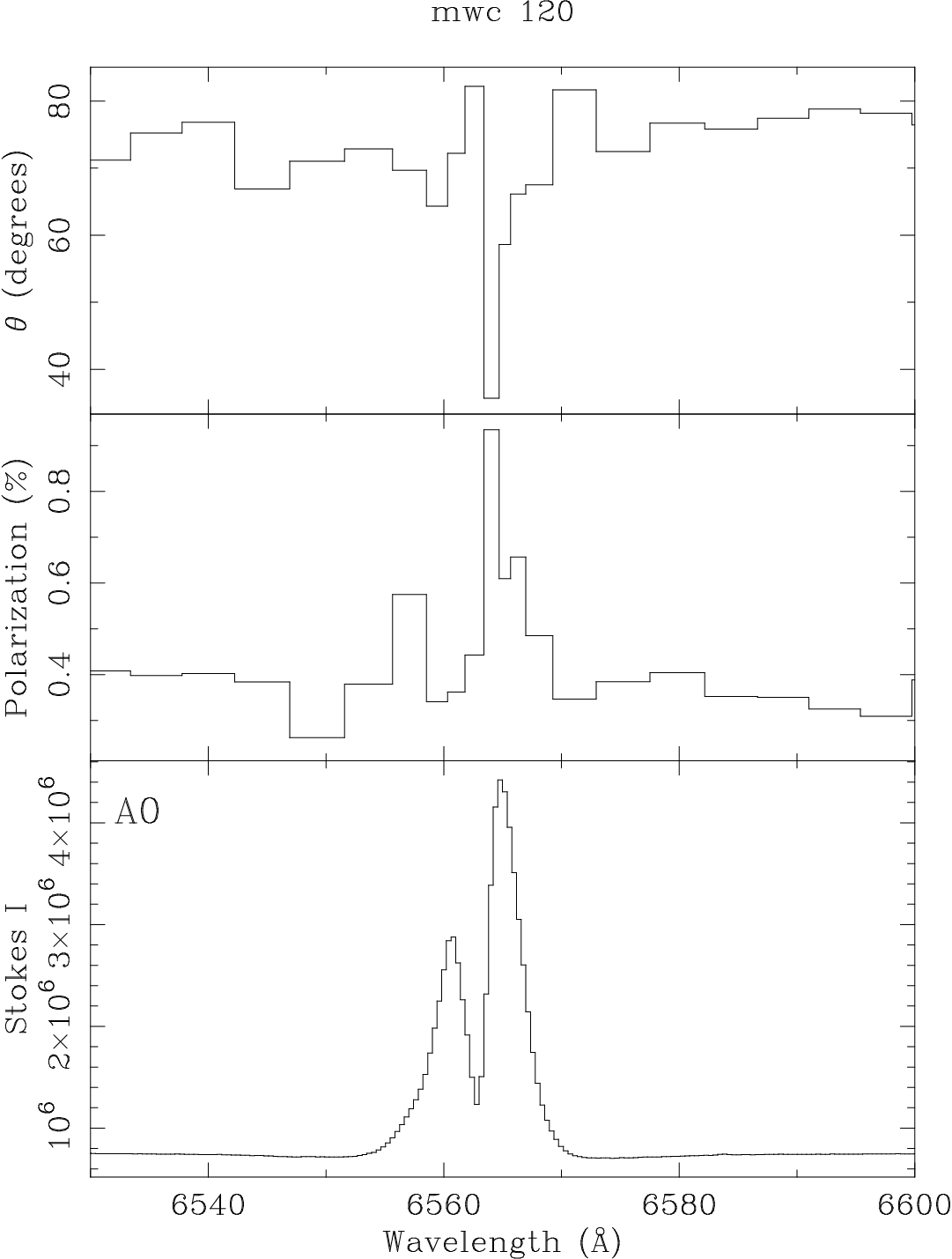}
\epsfxsize=0.26\textwidth\epsfbox{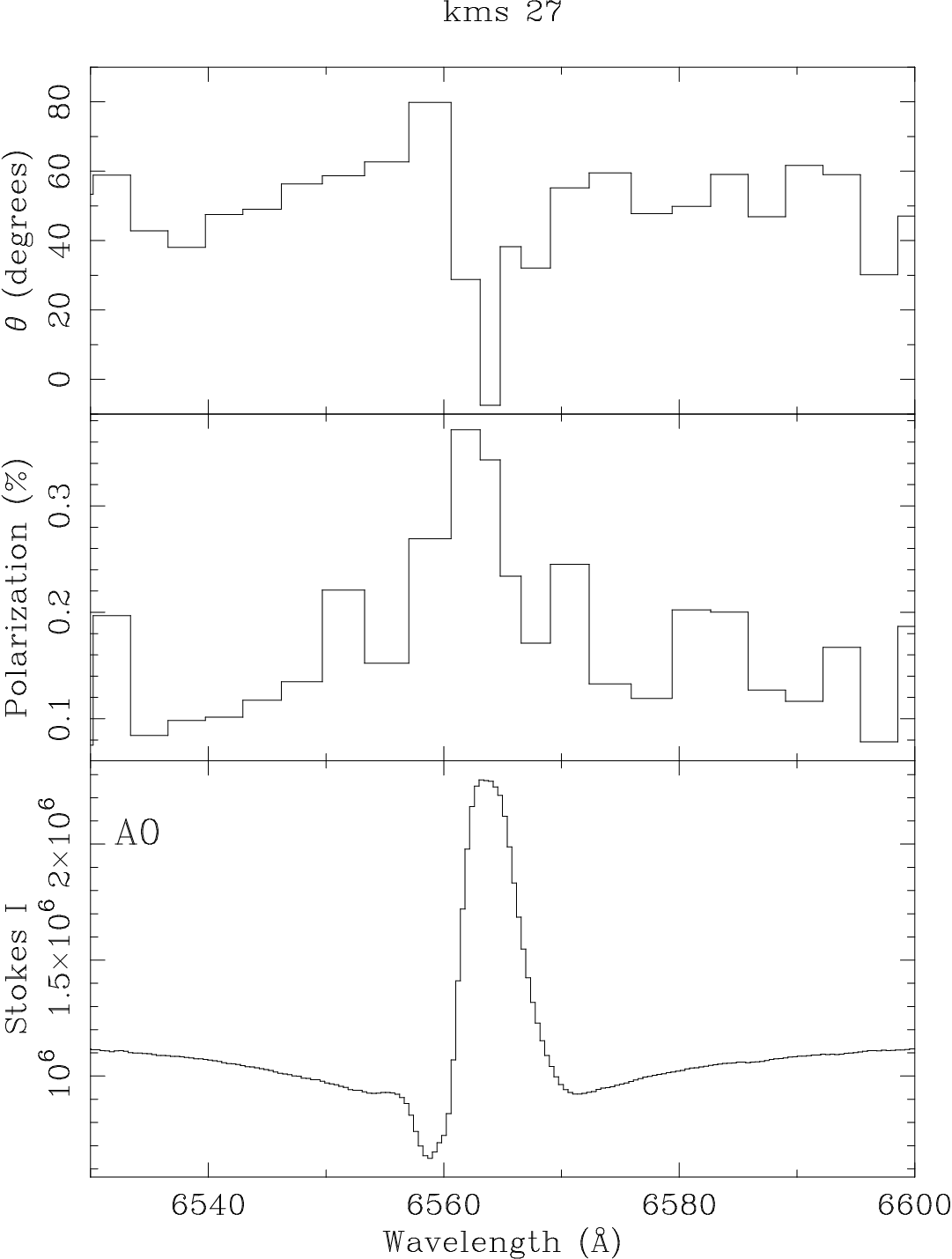}
\epsfxsize=0.26\textwidth\epsfbox{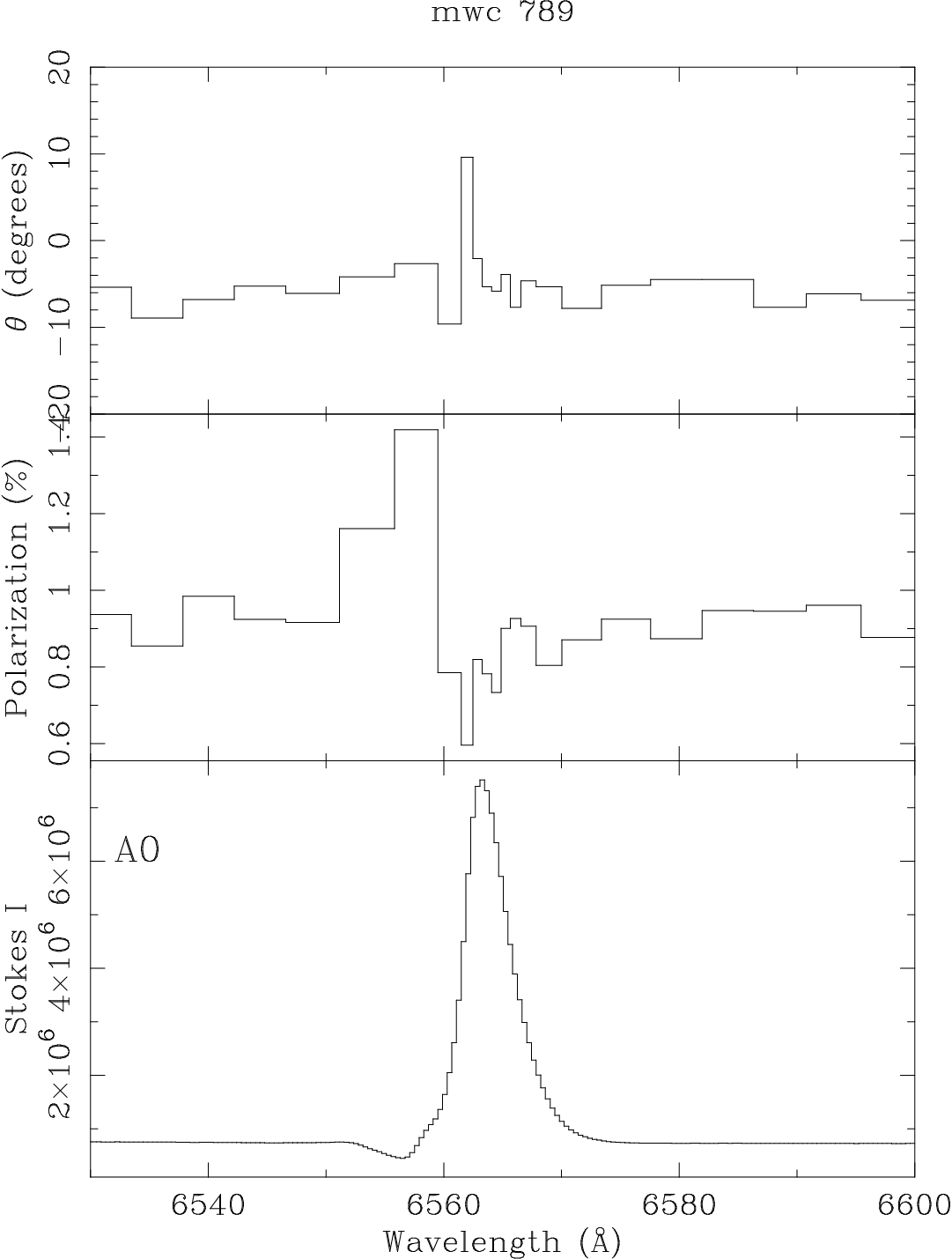}
\epsfxsize=0.26\textwidth\epsfbox{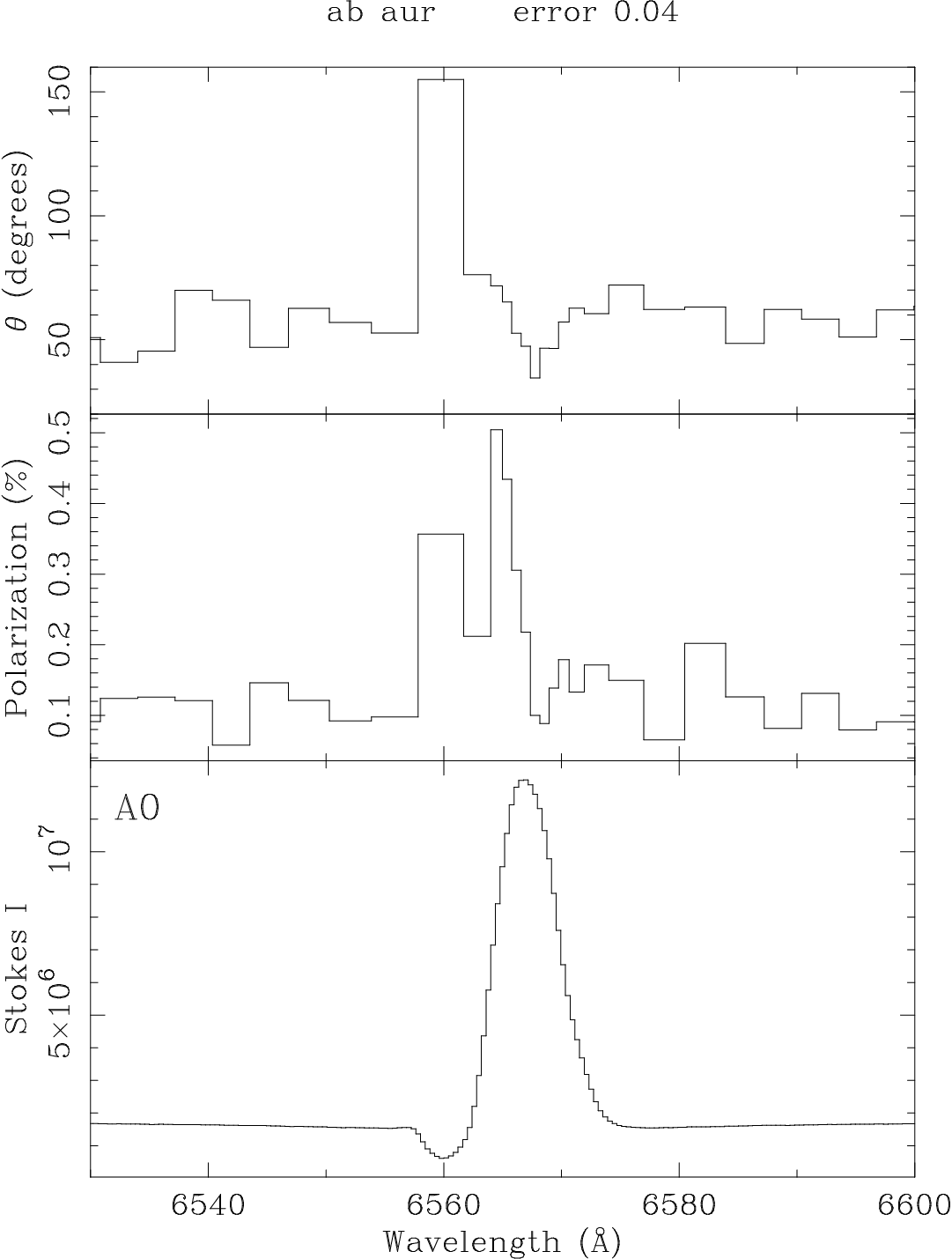}
}
\mbox{
\epsfxsize=0.26\textwidth\epsfbox{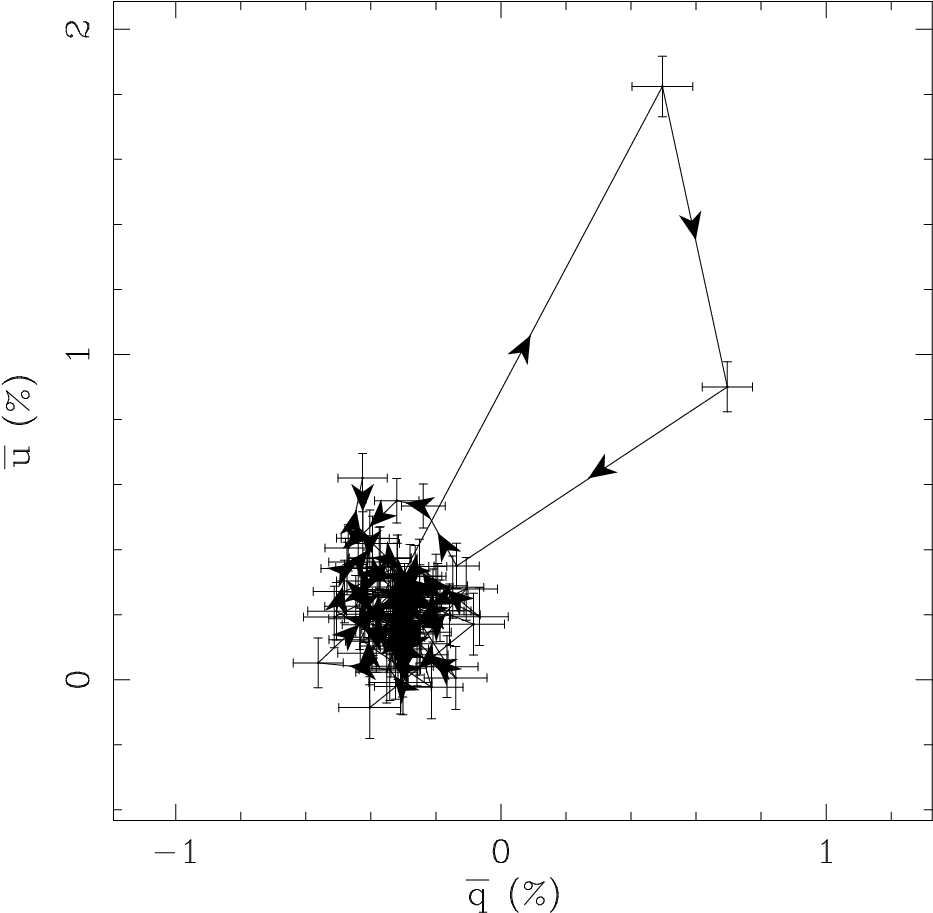}
\epsfxsize=0.26\textwidth\epsfbox{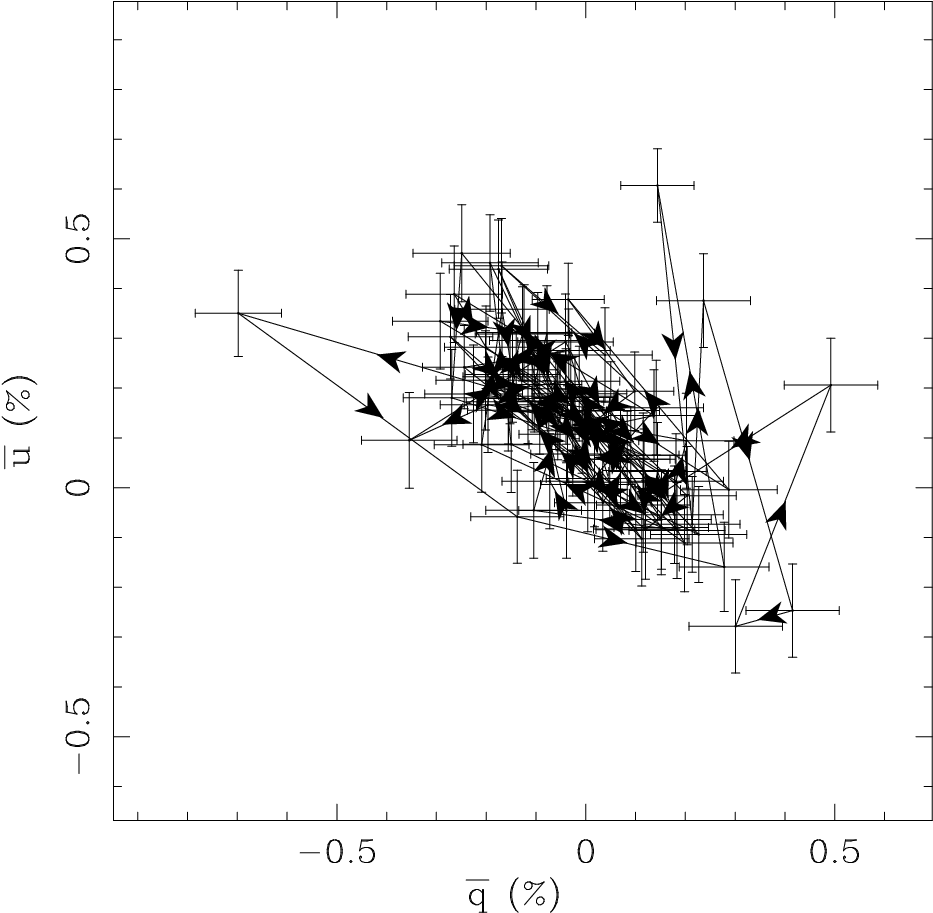}
\epsfxsize=0.26\textwidth\epsfbox{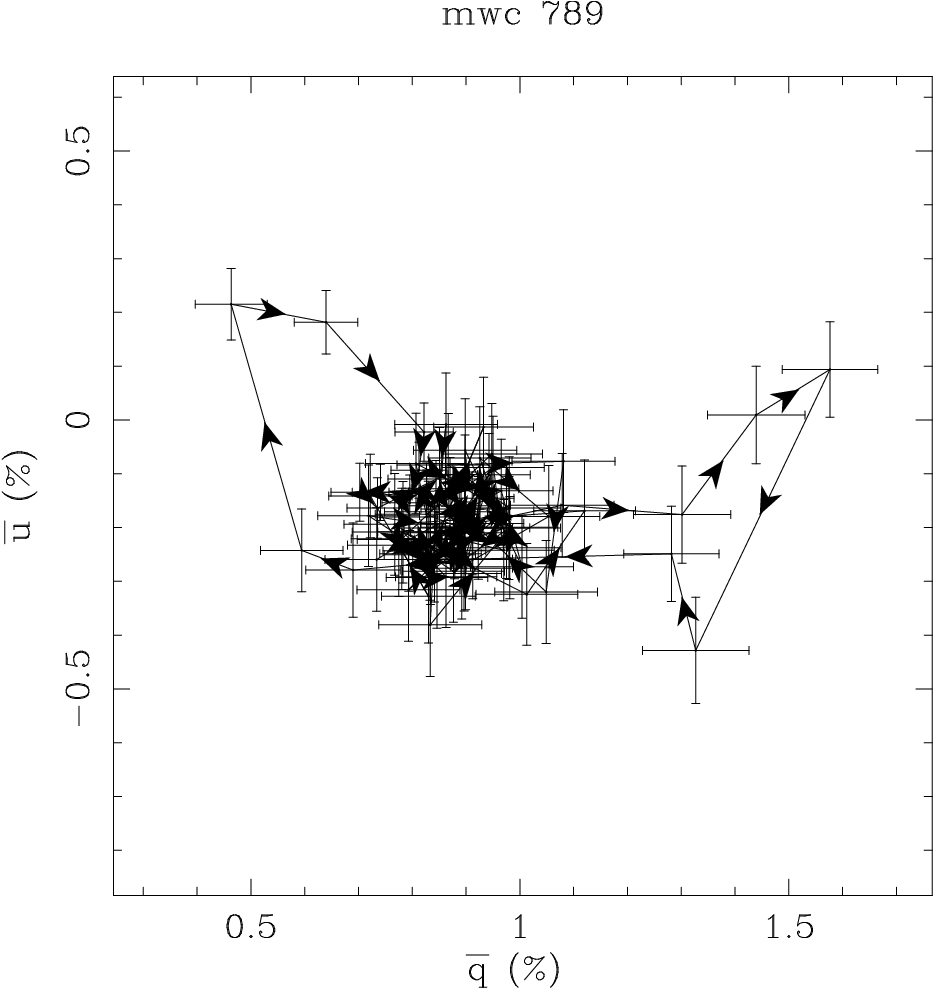}
\epsfxsize=0.26\textwidth\epsfbox{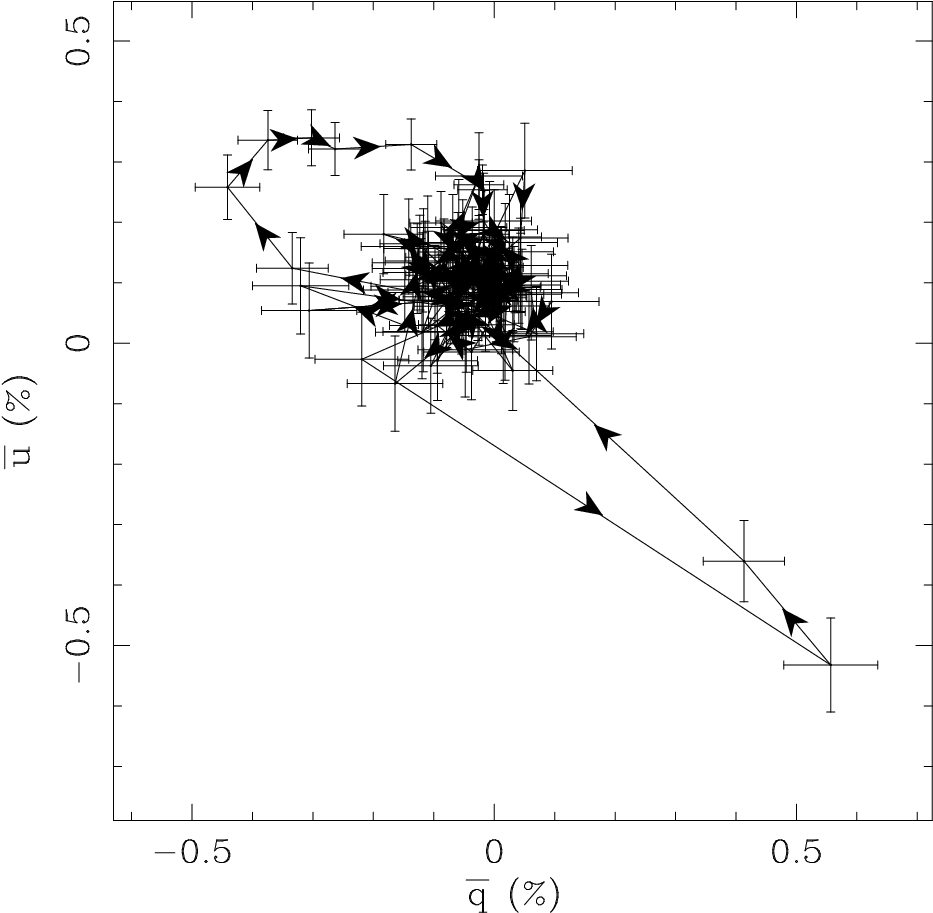}
}
\mbox{
\epsfxsize=0.26\textwidth\epsfbox{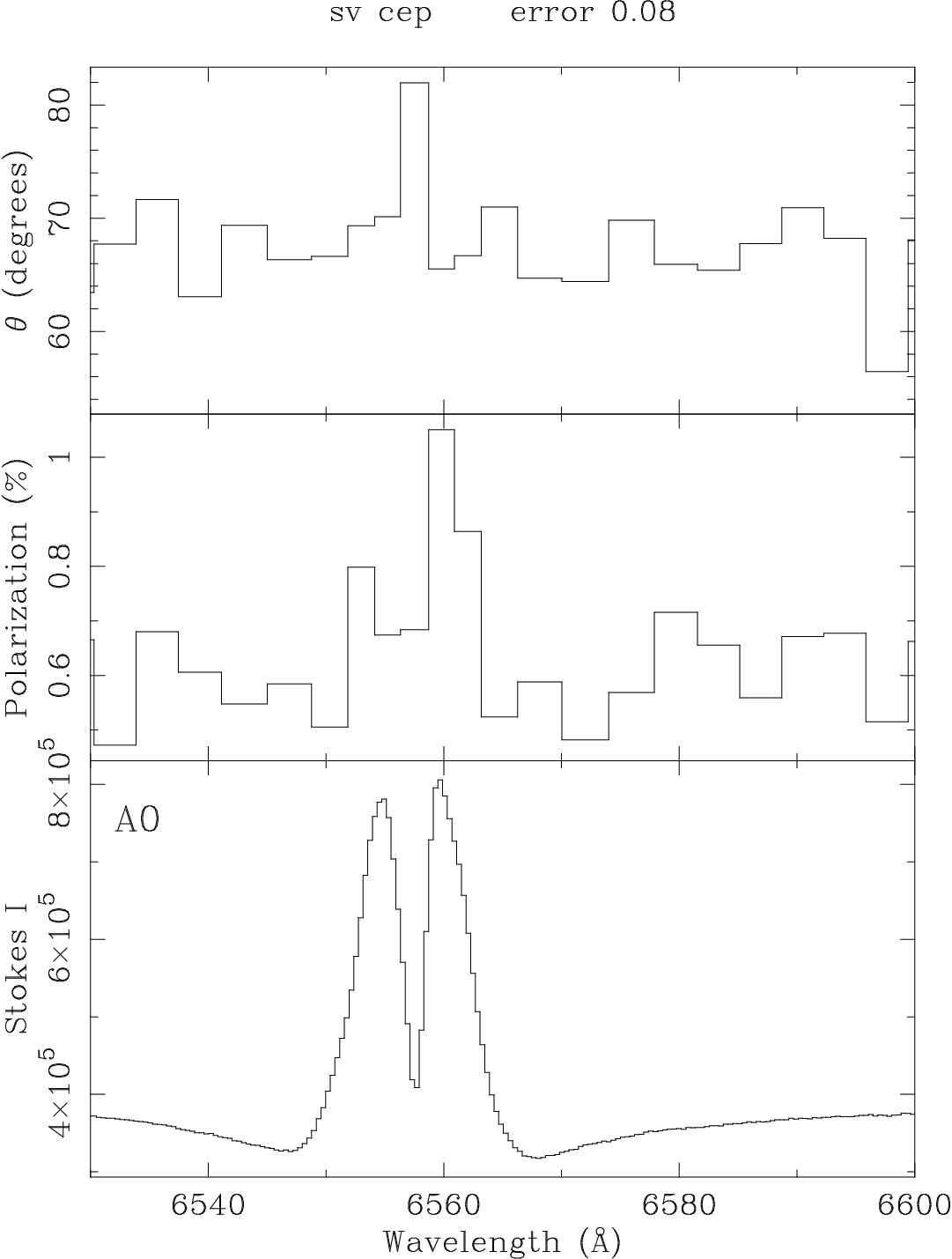}
\epsfxsize=0.26\textwidth\epsfbox{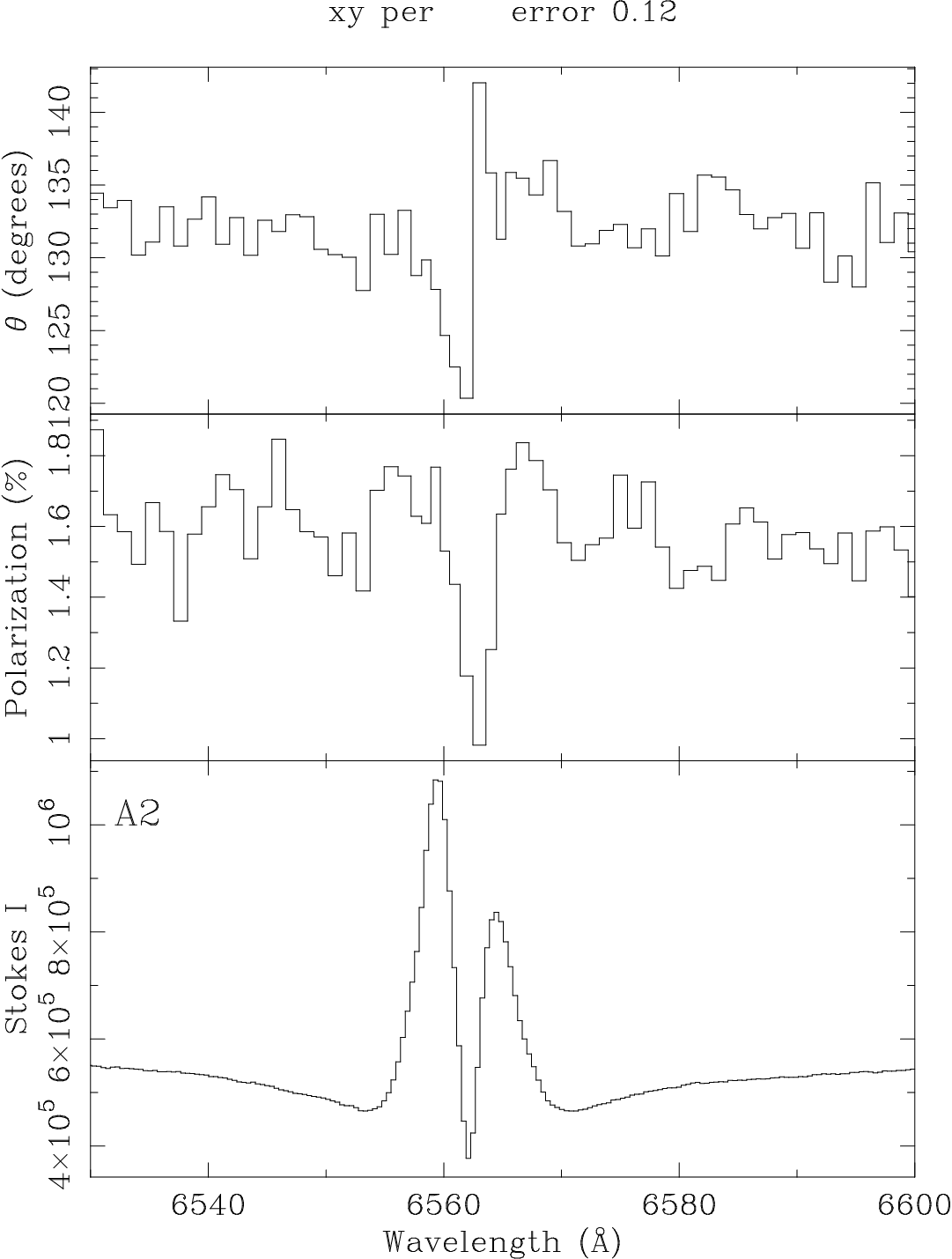}
\epsfxsize=0.26\textwidth\epsfbox{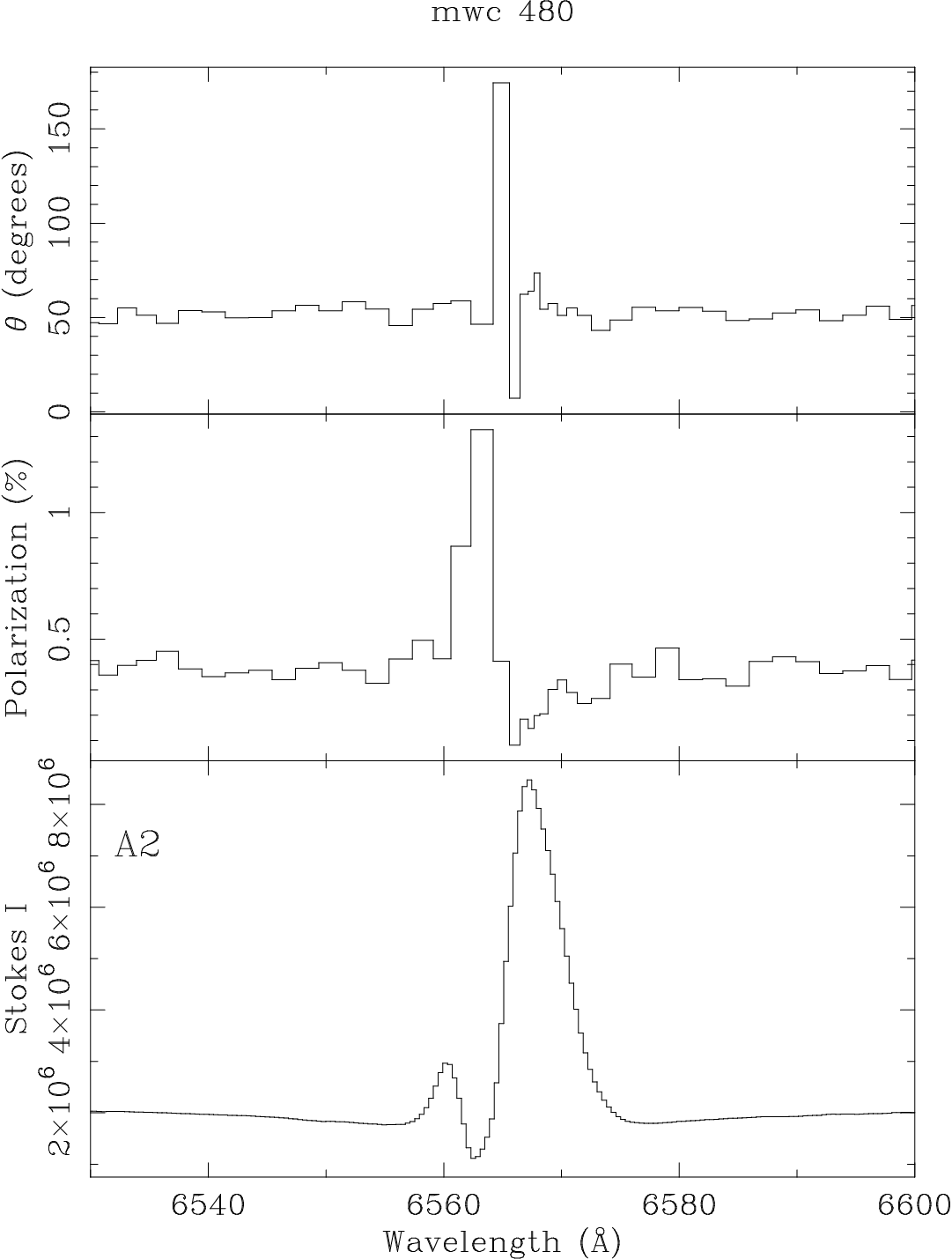}
\epsfxsize=0.26\textwidth\epsfbox{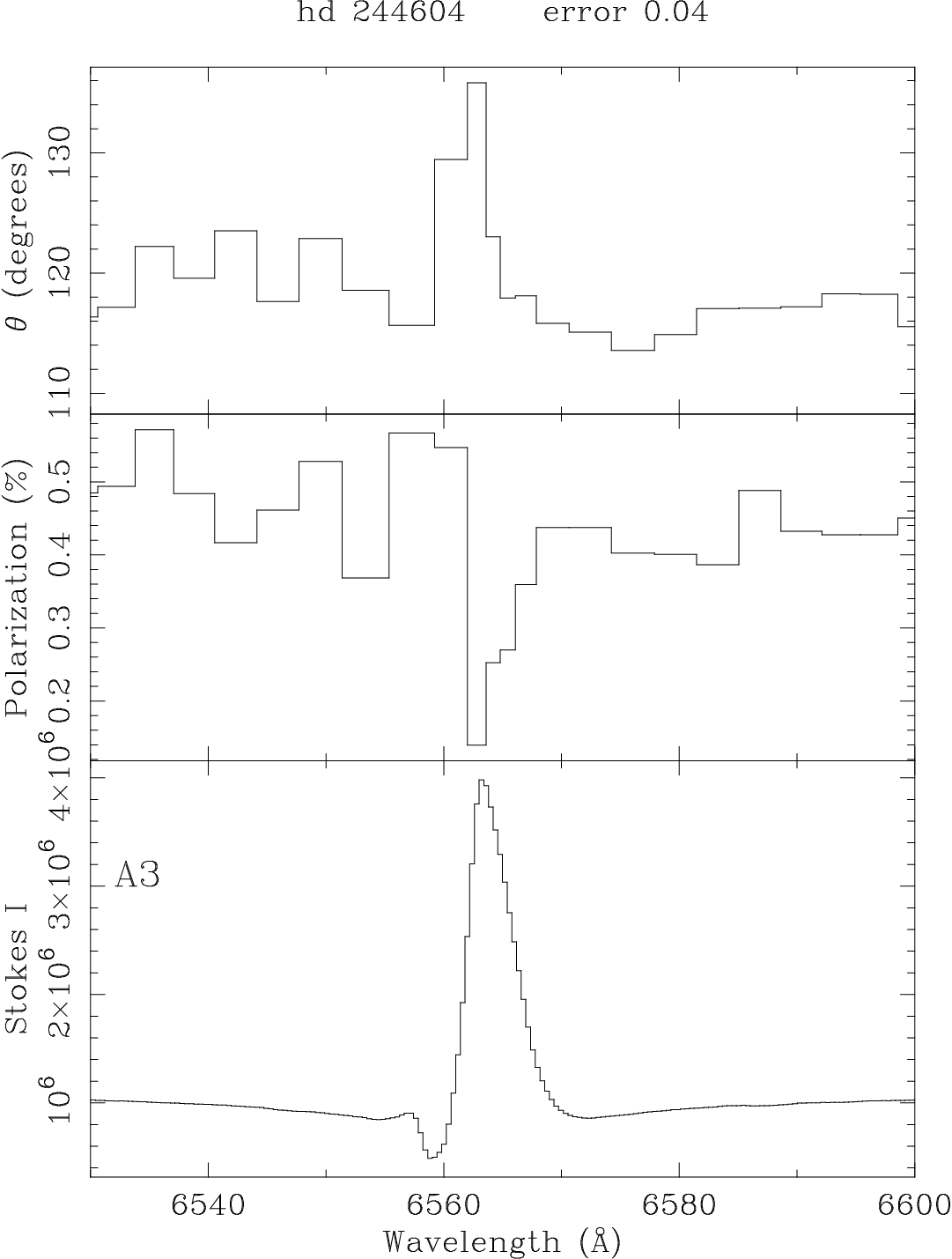}
}
\mbox{
\epsfxsize=0.26\textwidth\epsfbox{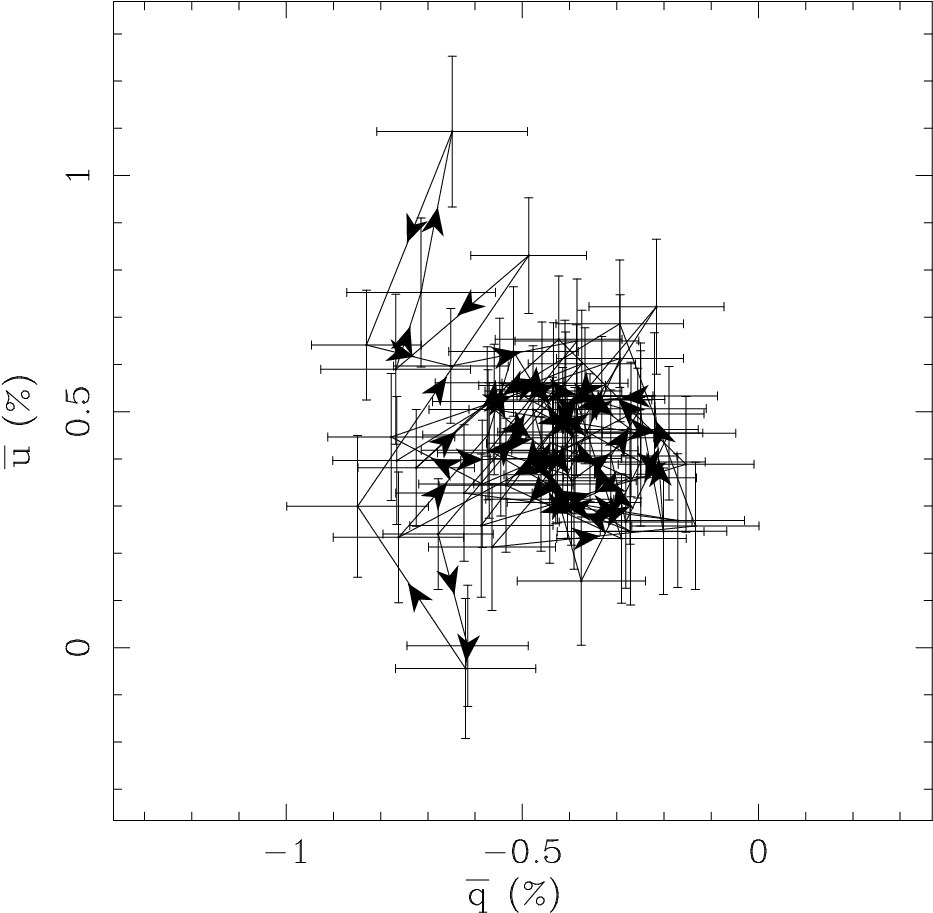}
\epsfxsize=0.26\textwidth\epsfbox{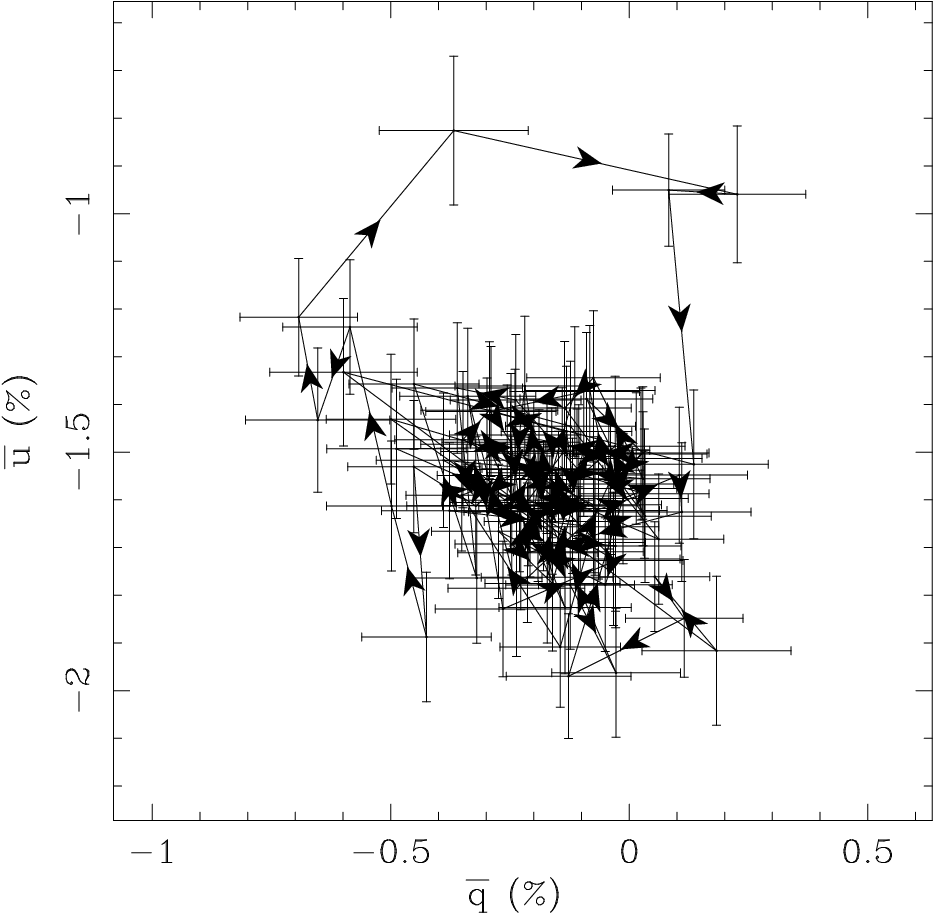}
\epsfxsize=0.26\textwidth\epsfbox{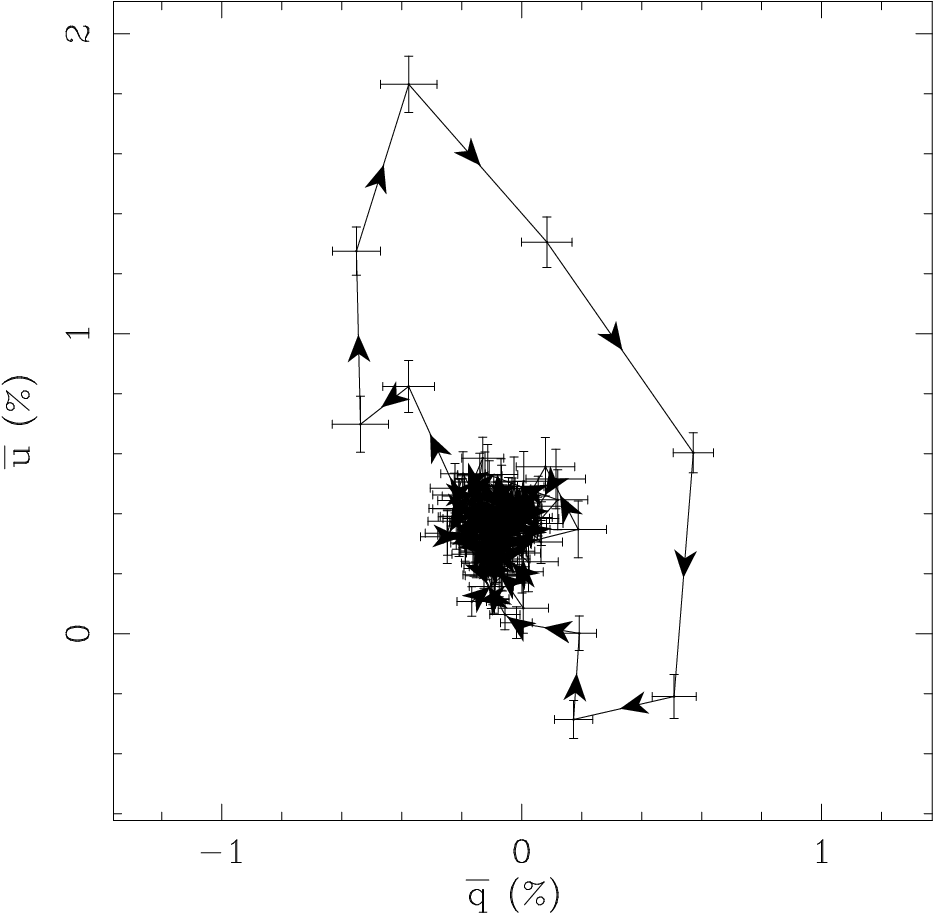}
\epsfxsize=0.26\textwidth\epsfbox{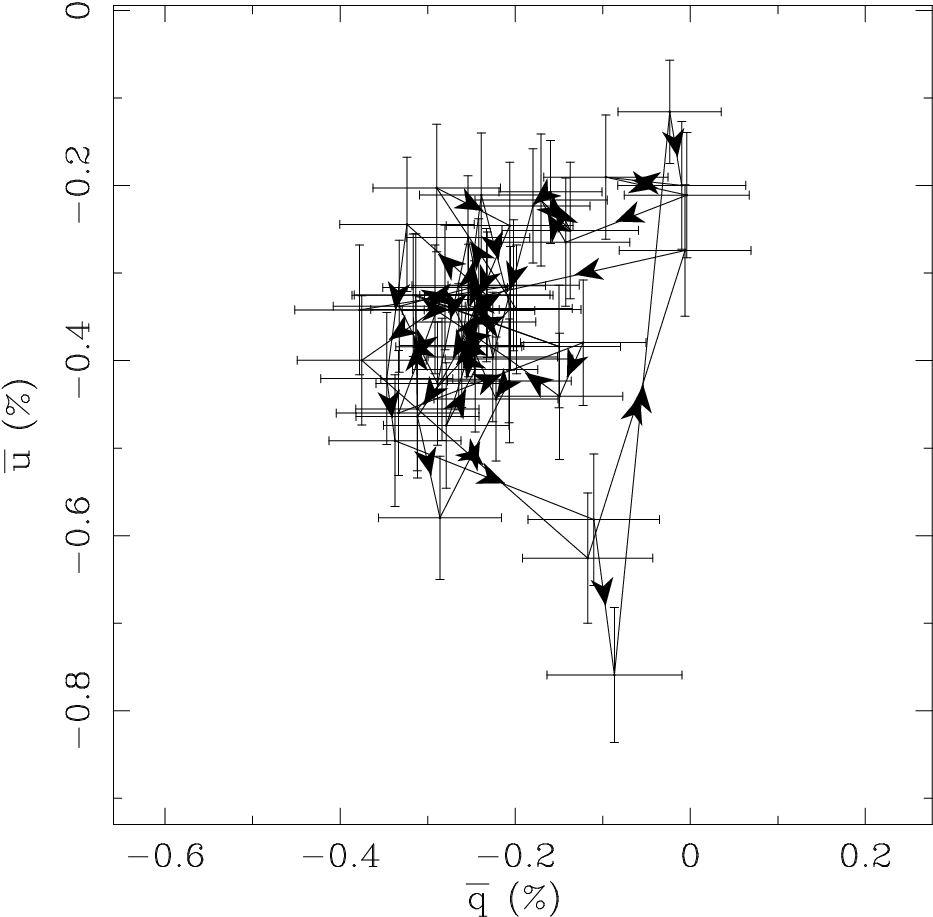}
}
\caption{Polarisation spectra of Herbig Ae stars and their 
$(Q,U)$ diagrams. See the caption of 
Fig.~\ref{f_hbe} for further details.
\label{f_hae}
}
\end{figure*}

\addtocounter{figure}{-1}%

\begin{figure*}
\mbox{
\epsfxsize=0.26\textwidth\epsfbox{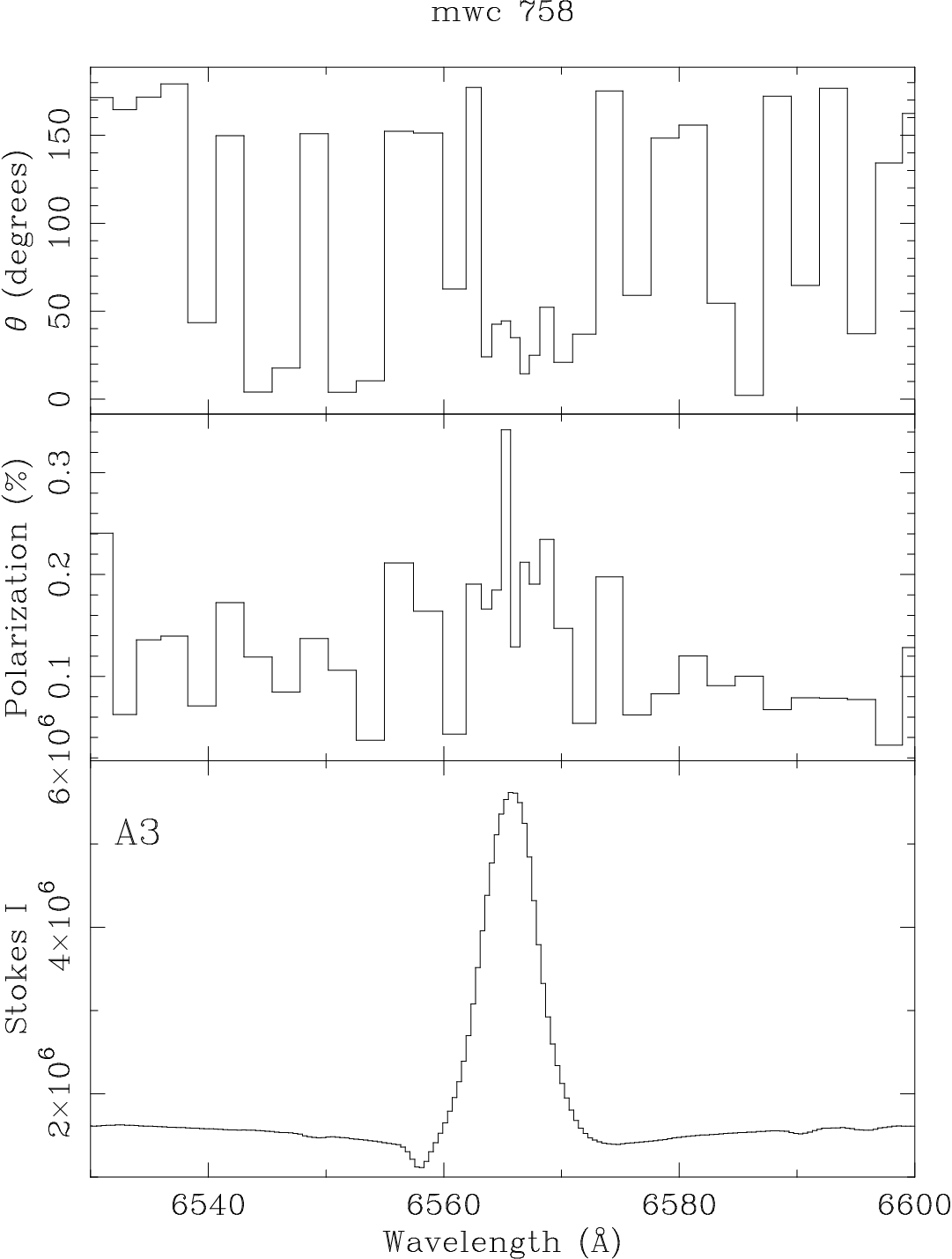}
\epsfxsize=0.26\textwidth\epsfbox{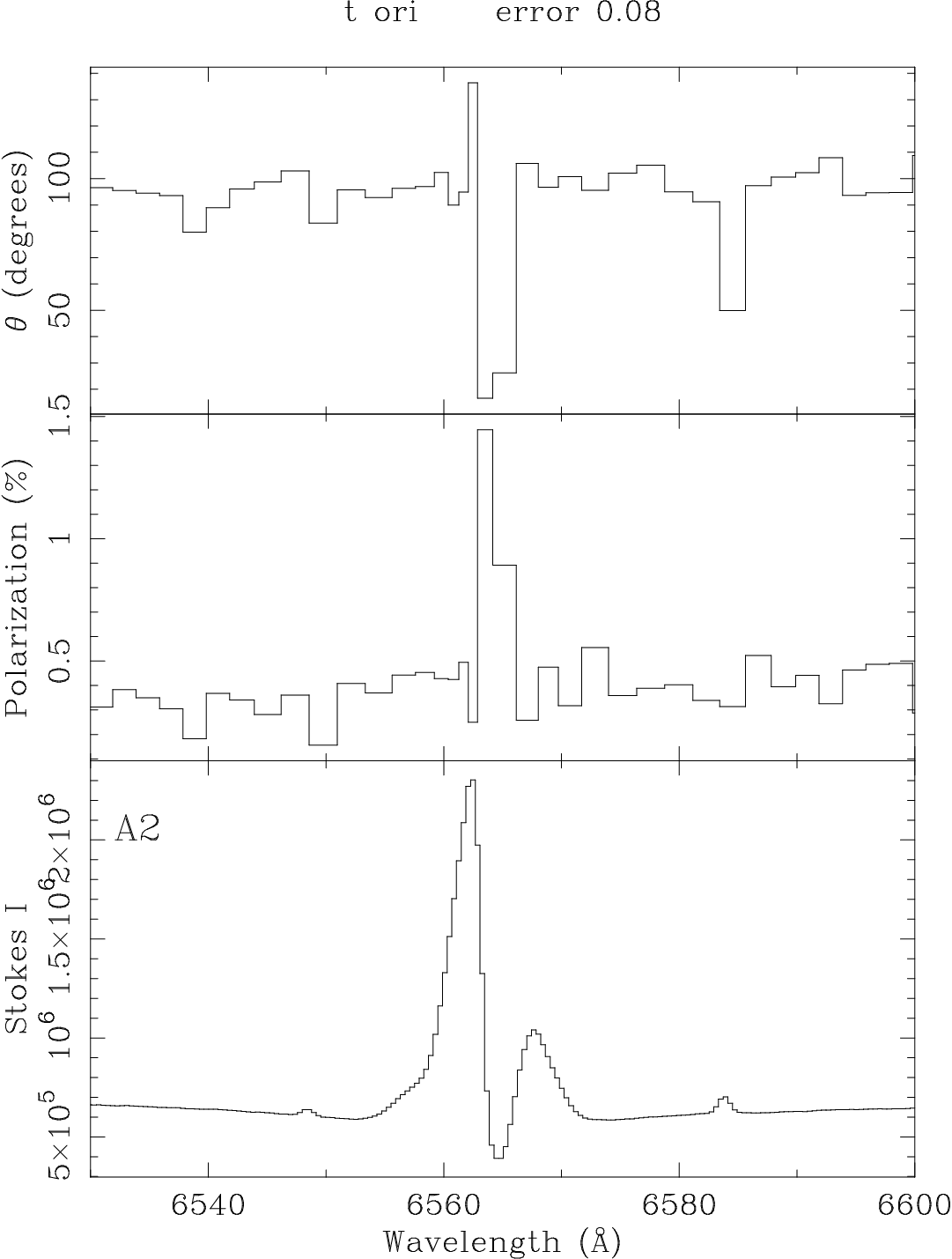}
\epsfxsize=0.26\textwidth\epsfbox{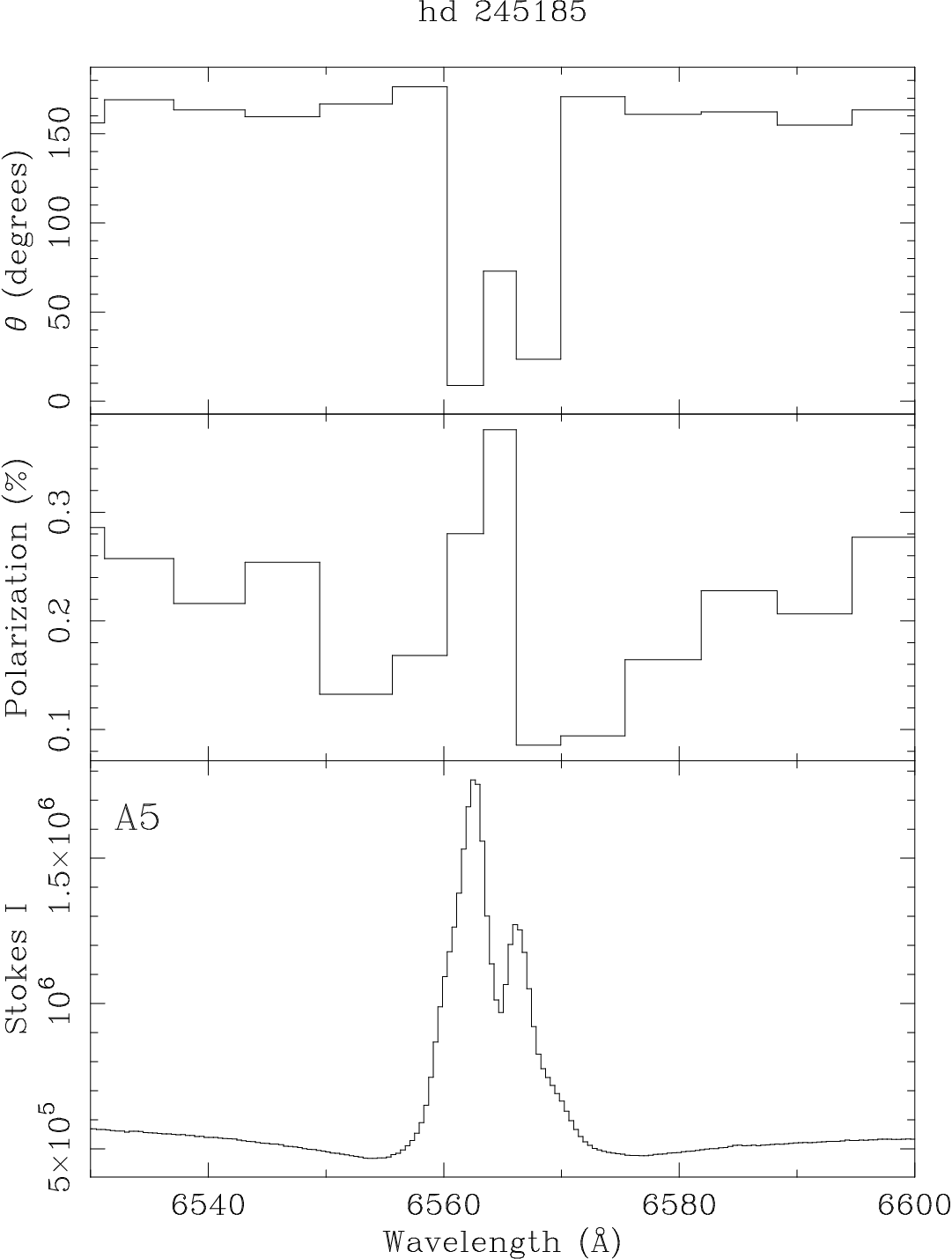}
\epsfxsize=0.26\textwidth\epsfbox{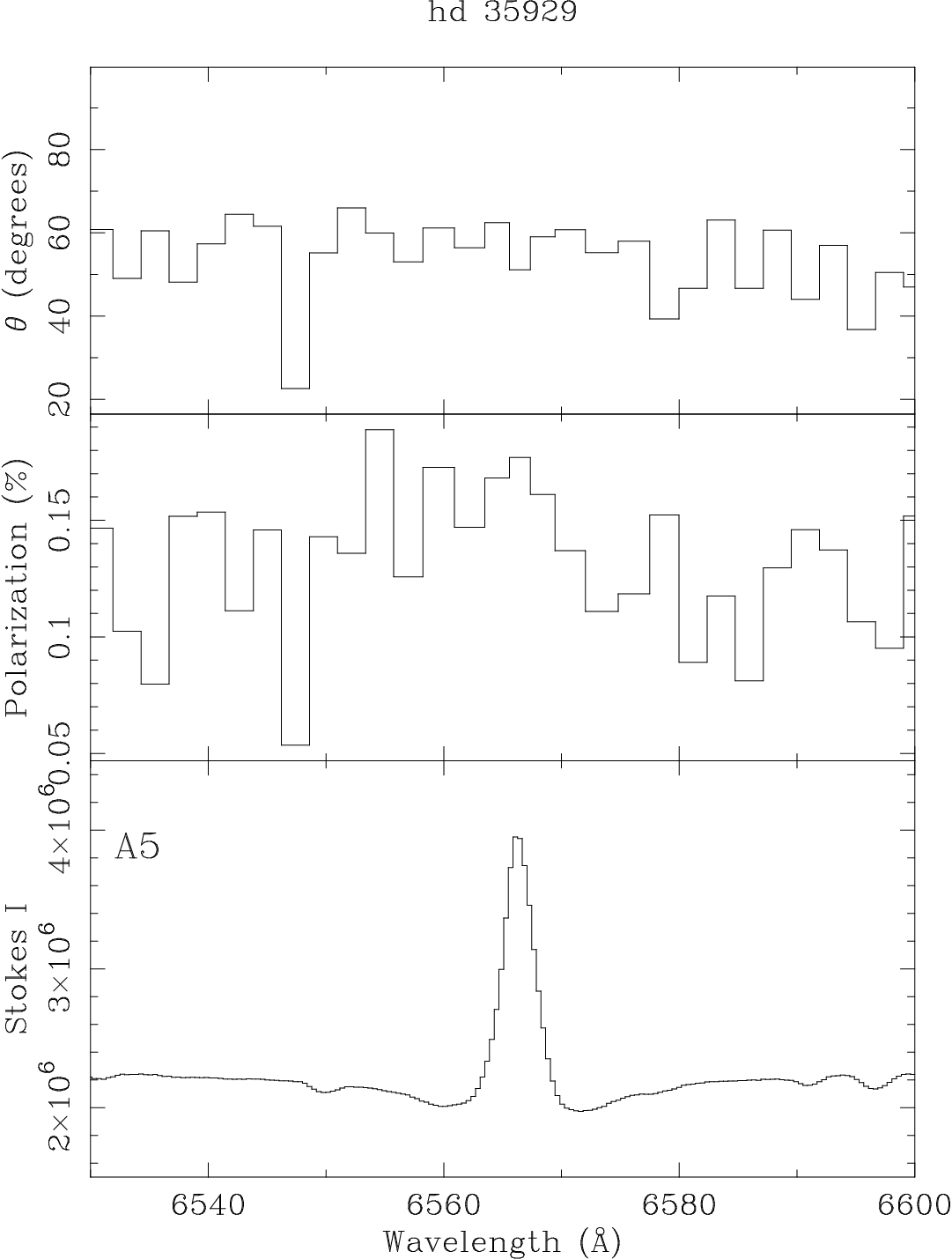}
}
\mbox{
\epsfxsize=0.26\textwidth\epsfbox{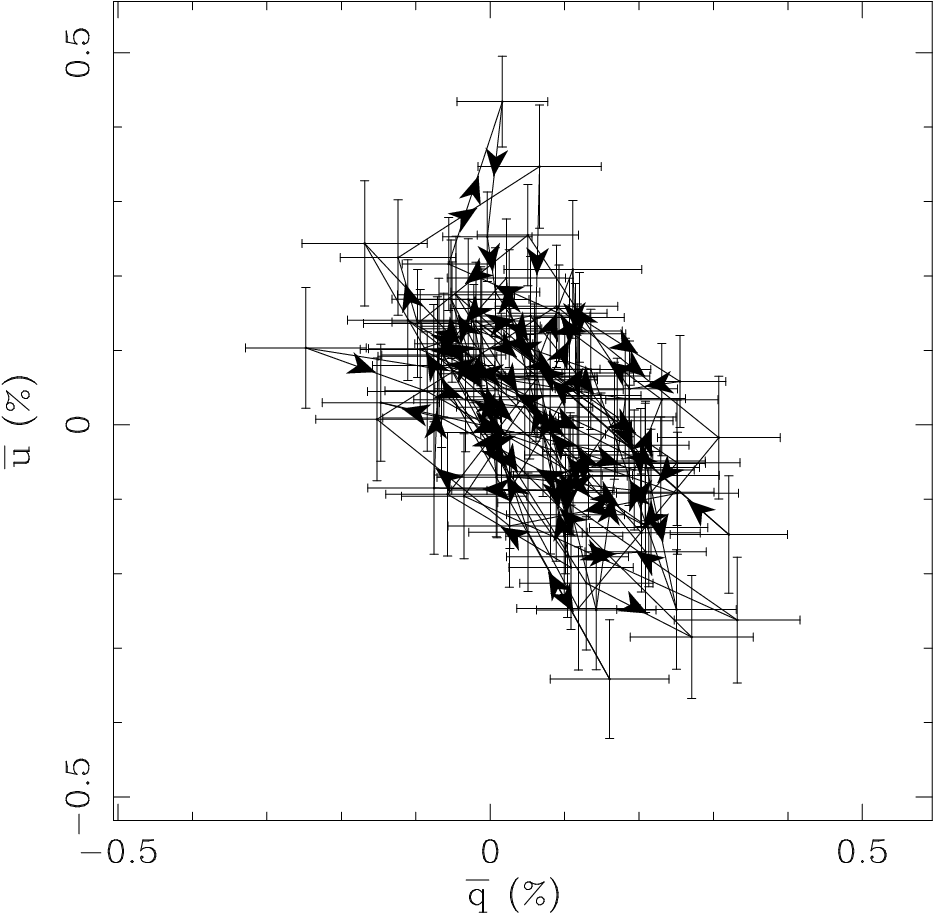}
\epsfxsize=0.26\textwidth\epsfbox{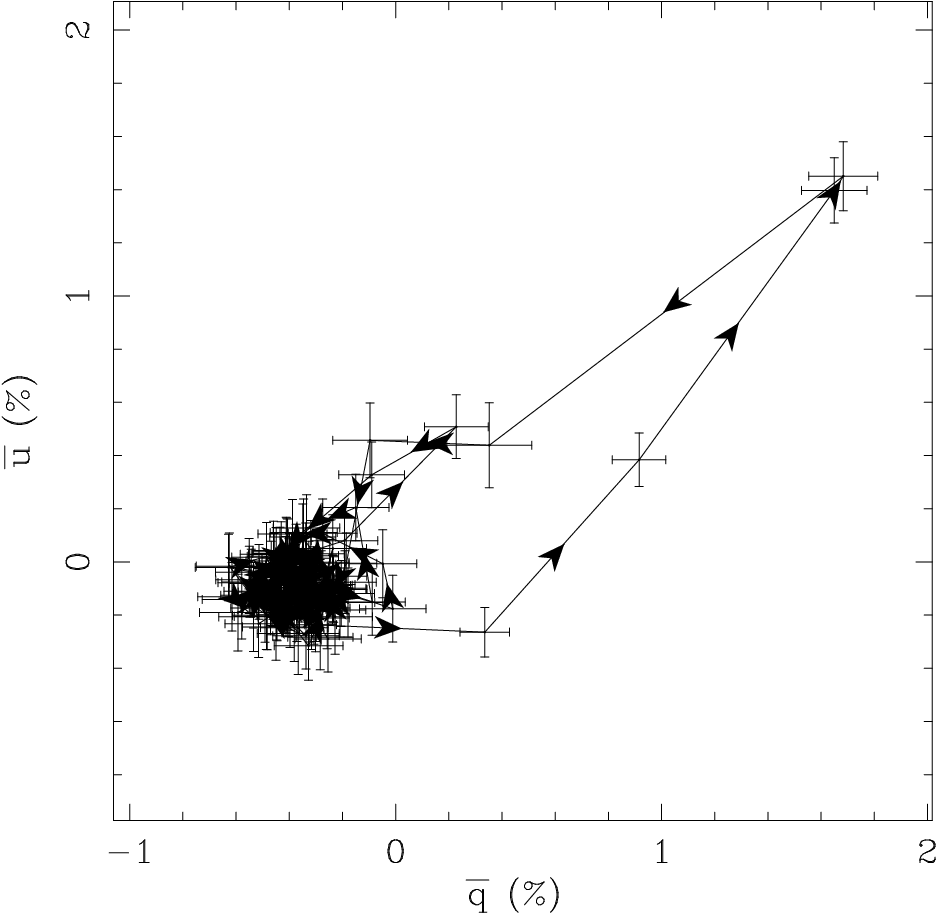}
\epsfxsize=0.26\textwidth\epsfbox{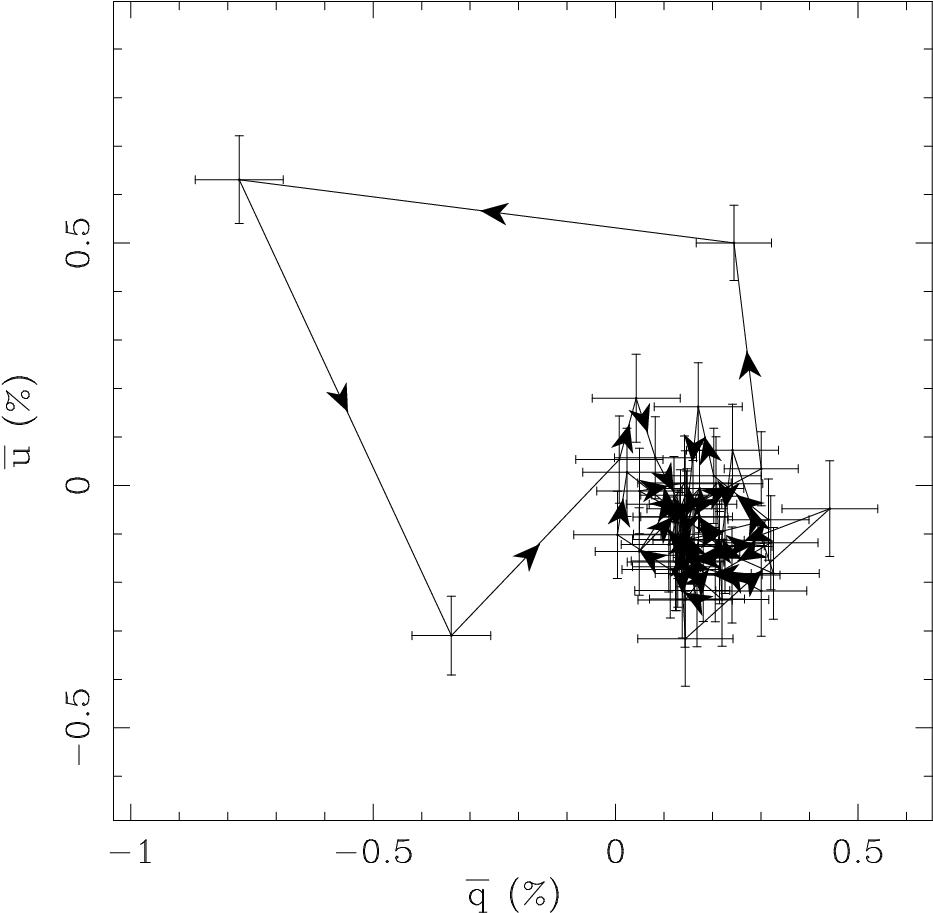}
\epsfxsize=0.26\textwidth\epsfbox{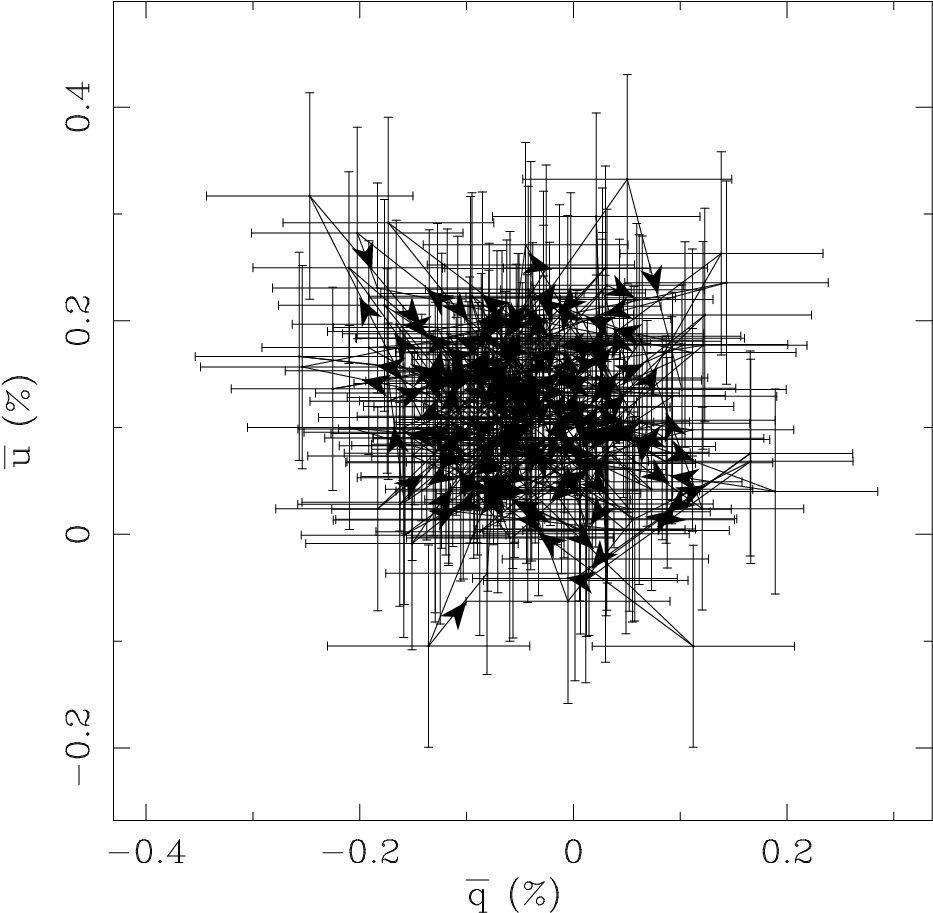}
}
\mbox{
\epsfxsize=0.26\textwidth\epsfbox{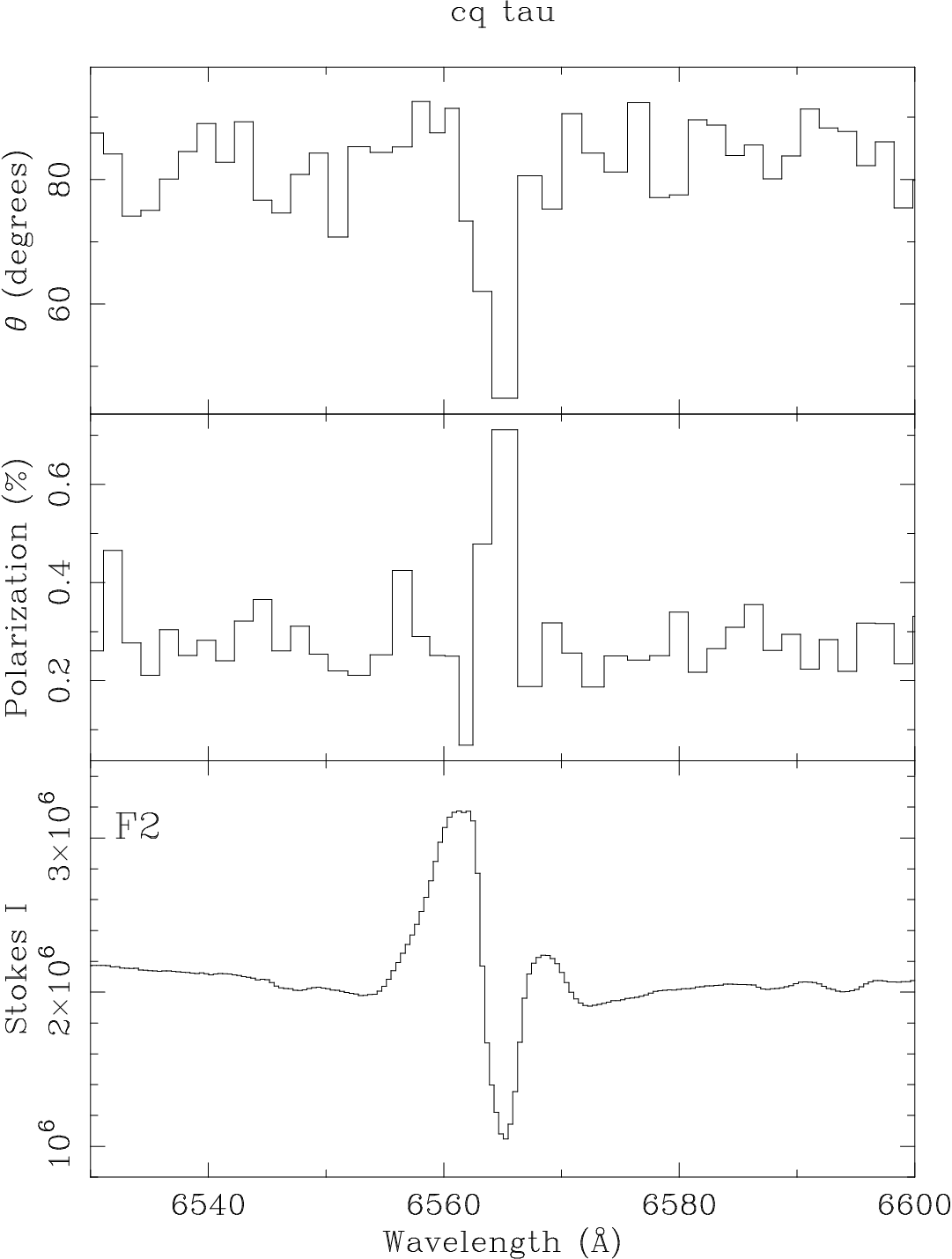}
}
\mbox{
\epsfxsize=0.26\textwidth\epsfbox{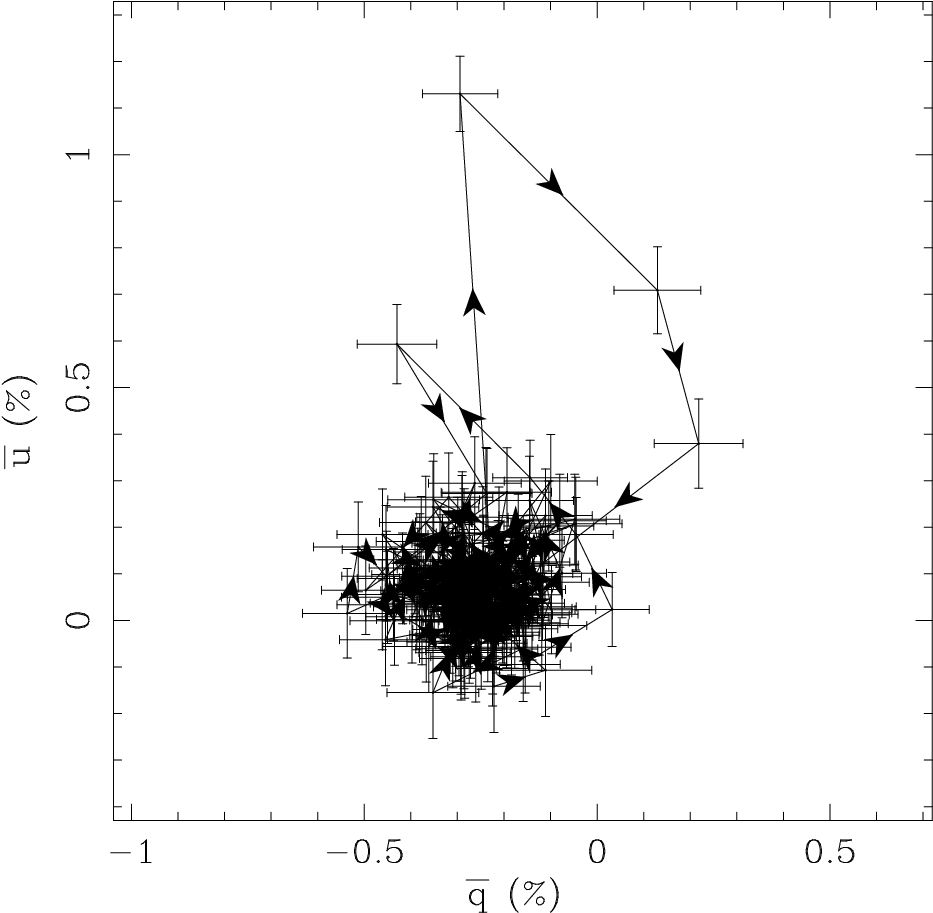}
}
\caption{Continued}
\label{f_hae2}
\end{figure*}

\subsubsection{Spectropolarimetric Tools}

Inspecting the different objects in Figs.~\ref{f_hbe} 
and~\ref{f_hae} one finds a range of morphology 
in the polarisation spectra (triplots), as well as in 
the $(Q,U)$ plots. Most importantly, 15 out of the 22 
Herbig stars show a clear variation across the \ha\ line. 
(The only objects that do definitely not show any changes across 
\ha\ at all in our data are IL~Cep, MWC~758, and HD~35929). The question 
whether or not a line effect is present or not is answered 
qualitatively in column (4) of Table~\ref{t_line}. The high number of 
line effect detections is interesting in its own right, given that 
any polarisation change across a line indicates scattering off a geometry, 
that, in projection on the sky, deviates from spherical symmetry.

Although the morphology in both the polarisation triplots 
and  $(Q,U)$ space shows some variety, we will classify 
them according to their common characteristics. To this end, we have 
developed several tools to characterise the line effects. One such tool 
involves the fractional line width over which the polarisation changes
(see data in columns 5 and 6 in Table~\ref{t_line}).
In cases where the widths of the PA and \%Pol changes are unequal, 
we choose the largest in specifying the change $\Delta 
\lambda(\rm Pol)$ (column 5).  This then is divided by 
the width of Stokes I, ${\Delta \lambda(\rm I)}$ to yield
a dimensionless measure of polarisation width (column 6).  Note that both
$\Delta \lambda(\rm Pol)$ and ${\Delta \lambda(\rm I)}$ are measured
at full width--zero intensity (FWZI).  

The other, more qualitative, tool concerns the {\it shape} of the 
polarisation change across the line (columns 7 and 8).  Whether or not the 
change in polarisation across \ha\ is consistent with the depolarisation 
mechanism is noted in column (7). To limit our subjectivity, we have 
constructed a recipe for answering this question that proceeds as follows. 
First, the line effect needs to be broad enough to be consistent with a 
depolarisation. If the ratio \ratpol\ is smaller than a value of 0.75 we 
claim that the line effect is not consistent with the depolarisation 
mechanism, and the answer in column (7) is simply a ``no''.  If however, 
the ratio \ratpol\ has a value larger than 0.75, the polarisation change 
could indeed be due to depolarisation, but not necessarily. We therefore employ 
a second criterion that takes into account the finer details of 
the line effects. In many cases a so-called ``flip'' in PA and/or \%Pol\ 
across the line will be clearly noticeable. In case this occurs the line 
effect can not be due to depolarisation and the answer in column 
(7) should also be ``no''.  If, on the other hand the behaviour in \%Pol\ 
and/or PA is monotonic, and \ratpol\ is larger than 0.75, the 
answer to the question ``Is the line effect due to depolarisation?'' can 
be answered convincingly with a ``yes''.  

As the shapes of the line polarisations are so nicely represented as 
loci in $(Q,U)$ space, we also classify 
the $(Q,U)$ plane morphologies.  These are given in column (8) of 
Table~\ref{t_line}.  In every case we expect there to be a dark knot of 
points that arises from the sampling of the continuum polarisation.  Where the morphology 
also includes a rounded loop the term ``loop'' appears in column (8)
of Table~\ref{t_line}. The word ``dot'' is given when no line effect is 
present, and finally in case there is a 
more or less linear excursion arising from the continuum knot, we 
refer to this excursion with ``exc''. For convenience, the spectral 
types of the objects are listed in column (9). 

\section{Individual Targets}
\label{s_ind}

First, we add to OD99's sample of Herbig Be stars (Sect.~\ref{s_hbe}), and 
then we turn attention to the later-type, and presumably less-massive Herbig Ae 
stars (Sect.~\ref{s_hae}).  A comparison between the 
spectropolarimetric signatures of HAe and HBe stars is made 
in Sect.~\ref{s_diff}.

\subsection{The Herbig Be stars}
\label{s_hbe}

The \ha\ line profile data obtained for the Herbig Be stars are presented in 
Fig.~\ref{f_hbe}.  Six out of eight HBe stars in Fig.~\ref{f_hbe}
show a clear line effect, whereas only two cases (IL~Cep and MWC~166)
do not show a line effect at all. Note that two stars (MWC~158 and AS~477) 
are not included in this statistics, for reasons discussed later on.
The size of this sample can be increased by including the data on the 
HBes from OD99.
The sample of Herbig Be stars in OD99 consists of nine HBe stars, of which four
stars are in common with the present study, which would leave us with 
five HBes to be added to the eight in the current study.
However the number of photons counts for Lk \ha\ 218 is regarded as insufficient and 
the object is thus dropped, leaving us with a total of (8+4) 12 HBes. As these four added 
HBes from OD99, namely V380~Ori, $\omega$~Ori, GU~Cma and MWC~297 are all cases 
of no discernible polarisation changes across the line, we would find that 6 out of 
12 Herbig B stars show a line effect. But, as the OD99 and the 
current data were obtained at different epochs, spectropolarimetric changes 
may be seen in the targets common to both studies.
We omit a discussion of such variability in this paper, but note that 
most of the objects in OD99 classified as having a line effect 
exhibit one here as well. Yet, there is an exception.
In the case of MWC~166 (HD~53367) the line effect noted by OD99 has 
disappeared. Although there is the possibility that what was once a 
detectably flattened circumstellar geometry has become more 
circular in projection onto the sky, it is more likely that the 
change is due to a reduced amount of ionised particles. Given that the 
new data show a large fall in the \ha\ equivalent width (now $-$2.8\AA\ rather 
than $-$14\AA\ as in OD99), it is likely that the 
reduced ionisation level resulted in fewer electron scatterings.
Adding MWC~166 as a line effect case, we finally find that 7 out of 12 HBe 
stars show a line effect.  
This incidence of line effects among the HBe objects may be compatible with 
a picture in which all HBes are embedded in flattened geometries on small 
scales. The explanation for the non-detections is then simply the consequence 
of an expected number of close to face-on-viewing angles, liable to produce 
too weak a polarisation to be detected. 
 
Let us briefly discuss the polarisation spectra (and $(Q,U)$ diagrams) 
of all the Herbig Be stars, as  displayed in the different 
panels in Fig.~\ref{f_hbe}. 

\noindent {\bf Objects showing a depolarisation:}~{\it MWC~137, MWC~1080, 
BD+40 4124 and MWC~361}. These are all objects where at \ha\, at least one of 
the two polarisation signatures -- as presented by \%Pol and/or PA -- 
follows the shape of the Stokes I line profile monotonically.  The width 
over which the polarisation change occurs is also relatively broad, resulting 
in a ratio \ratpol\ larger than 0.75.  Therefore these objects may indeed be 
interpreted as undergoing a line depolarisation associated with  
continuum polarisation caused by scattering off an ionised disc on small 
spatial scales.  In the case of MWC~137, with its exceptionally high 
line/continuum contrast, the ripple correction (derived from the continuum) 
may not have been good enough to remove all ripple from the polarisation
spectrum, as seems to be the case near line centre in both the \%Pol and PA 
spectra.  This deficiency is responsible for the chaotic appearance of
the $(Q,U)$ plane excursion.  Yet, as the shape of the PA clearly follows the 
Stokes I line, the object can surely be regarded as being subject to 
depolarisation. The \%Pol in the spectrum of MWC~1080 follows the shape of 
the Stokes I line profile, and the object can thus also be considered as an 
example of depolarisation. For this object the $(Q,U)$ diagram does not yield 
much additional information because of the smear of points due to the changing
continuum polarisation. 
For BD+40~4124 both the PA and the \%Pol follow the intensity profile, 
consistent with the depolarisation picture. Although the \%Pol of 
MWC~361 shows a classical depolarisation across the line, the change in 
PA is rather less straightforward. Specifically, the blue wing does not 
show any sign of a change in PA, while the innermost red wing shows a flip 
and then a return to the local continuum PA in the far wing. We come back 
to this unusual case in the discussion in Sect.~\ref{s_disc}.

\noindent {\bf Objects showing a complex line effect:}~{\it MWC~147 and HD~58647}.
The late B type objects MWC~147 and HD~58647 both show polarisation 
changes wherein \ratpol\ is significantly smaller than 1.  Also, in 
the $(Q,U)$ plane the polarisation changes across the line do not present 
as simple linear excursions. 

\noindent {\bf Objects showing no clear line effect:}~{\it IL~Cep, MWC~166, AS~477}.
In a few cases we must classify the B star as not showing 
a line effect. In the case of IL~Cep this is unambiguous, but 
due to the low signal for AS~477 the data are 
not convincing one way or the other. Therefore we do not regard 
this object in our statistics and place a question mark in 
column (4) against this star in Table~\ref{t_line}.  Note that 
for MWC~166 one may possibly believe that there is a residual 
polarisation change across the line, but we view
the change as too narrow to be credible.\\

\subsection{The Herbig Ae group}
\label{s_hae}

The spectropolarimetric data for the HAes are presented in 
Fig.~\ref{f_hae}. The first thing that is striking is the -- perhaps 
unexpected -- high number of line effects, namely in 9 out of the 11 targets
for which the data quality allows sufficient definition. 
This is even higher than for the HBe sample. Furthermore, there seems to be a 
difference in the character of the polarisation changes compared to those 
seen in the HBe stars.  As for the HBe stars, we now present our
classification of these results.

\noindent {\bf Objects showing a loop:}~{\it MWC~120, XY~Per, MWC 480, T~Ori, 
HD~245185, and CQ Tau}. There are certainly 4 instances of very well 
formed loops in the $(Q,U)$ 
plane.  In the triplots these correspond to the line-centre flips (and
neighbouring points) in either \%Pol or PA. MWC~480, a relatively bright
HAe star, is a particularly fine example. Although the loops associated 
with T~Ori and MWC~120 are sufficiently elongated that they might 
alternatively be described as more akin to excursions, the complexity of 
the changes apparent in the triplots strengthens the similarity with the 
clear-cut examples of ``loops'' -- hence their inclusion here.\\  

\noindent {\bf Objects showing a loop and excursion:}~{\it MWC~789 and AB Aur}.
Two stars in the HAe groups show a behaviour in $(Q,U)$ space that is 
more complex than the presence of just a loop. These cases are 
MWC~789 and AB~Aur. Both are objects where the \ha\ line profile
includes a blueshifted absorption component: it is this part of the
line profile that maps onto the $(Q,U)$ plane as a more linear 
excursion.  The rounded loop in both is then associated with the
redshifted line emission. A fuller discussion of this morphology
is given in Pontefract et al. (2000).\\ 

\noindent {\bf Objects showing no line effect:}~{\it MWC~758 and HD~39529}.
In these two cases the data are good enough that we can claim no 
evidence of any kind of line effect.\\

\noindent {\bf Objects in need of better observations:}~{\it KMS~27 and SV~Cep}.
For these two objects the quality of the data is relatively poor and 
hence their interpretation is more difficult. 
The data on SV~Cep suffer from a too low signal, 
just as was the case for AS~477 in the HBe group.
KMS~27 suffers from a ``smear'' in $(Q,U)$ space, which is due to the broad 
photospheric absorption wings.
In view of the above, the objects KMS~27 and SV~Cep are dropped 
from our statistical overview.\\

\section{Trends and differences among the Herbig Ae/Be stars}
\label{s_diff}

The motivation behind this observational study is to discover if there
is any dependence of the geometry of the near-stellar environment of
Herbig Ae/Be stars upon spectral type.  

First we restate the raw statistics of the frequency of detectable 
linear polarisation changes across \ha\ among the Herbig Be
stars, as compared with the Herbig Aes.  We found that about half of
the HBe stars had detectable changes across the line (7 out of 12 to be 
precise), while this ``success rate'' rises to a more impressive 9 out 
of 11 among the HAe stars. Note that the present sample sizes are not yet
large enough for this difference to be formally significant.

\begin{figure} 
\mbox{\epsfxsize=0.48\textwidth\epsfbox{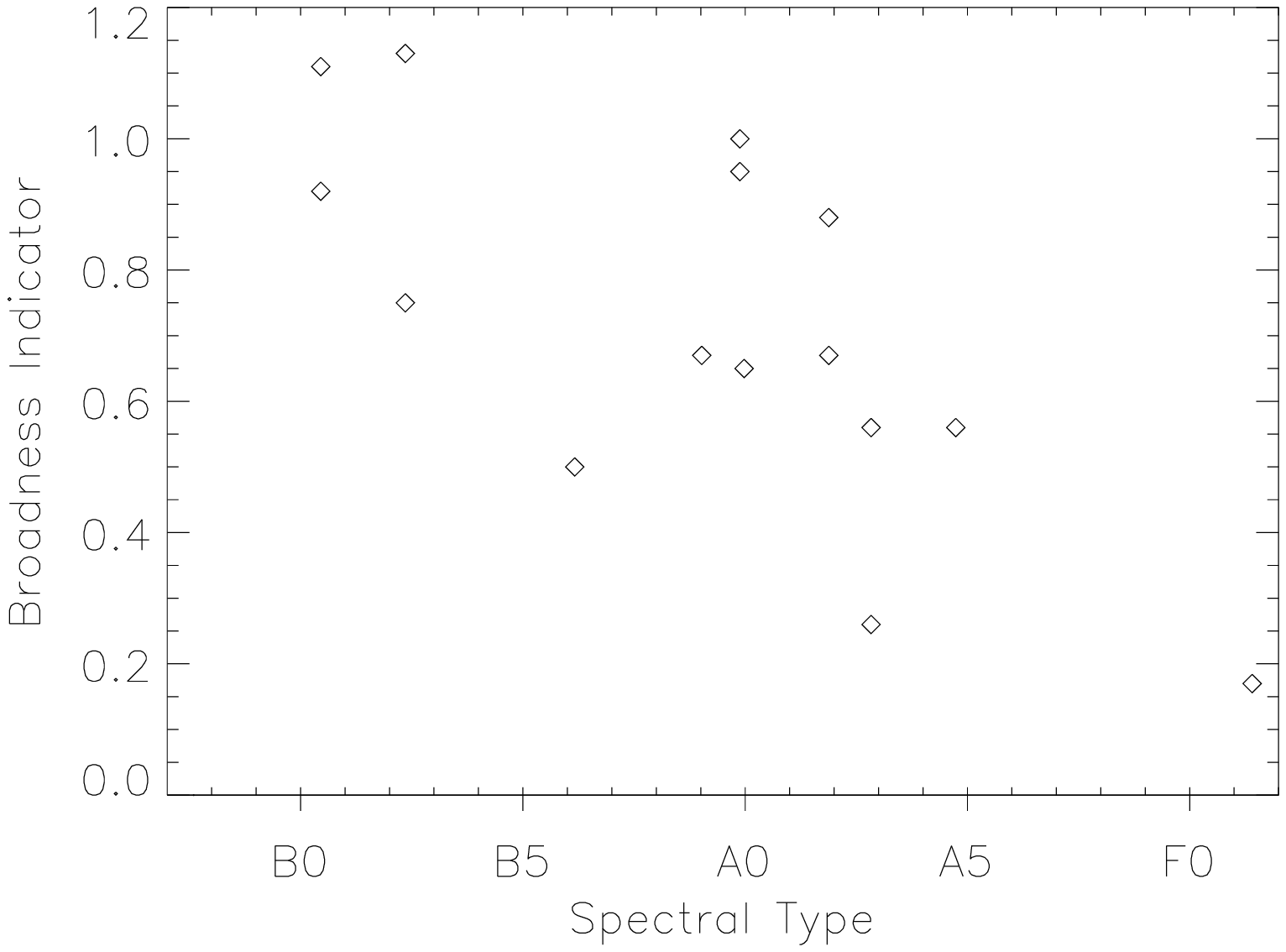}}
\caption{The fractional width \ratpol\ plotted against spectral 
type. The errors on the broadness indicator \ratpol\ are about 
10 per cent.}
\label{f_trend}
\end{figure}

A second attribute that shows some dependence on spectral type is the 
fractional width of the \ha\ line profile over which the polarimetric 
change occurs, specifically the quantity 
($\frac{\Delta \lambda(\rm Pol)}{\Delta \lambda(I)}$). These fractional 
widths -- as listed in column (6) of Table~\ref{t_line} -- tend to 
decrease toward later spectral type.  To visualise this, we have 
plotted the fractional widths in 
Fig.~\ref{f_trend}.  There is much scatter, in part due to the difficulty 
of
measurement, but presumably also due to real physical scatter.
Nevertheless, it is apparent that the fractional widths for e.g. the B0-B2
HBe stars tend to be higher than those observed at spectral types later
than A0: the means for these two sub-groups are respectively \ratpol\ $=$ 
1.0 $\pm$ 0.2 and 0.5 $\pm$ 0.3.

Of greater significance is that the Herbig Ae and Be stars present 
distinct 
kinds of \ha\ signature.  This can be seen by comparing Figures 1 and 
2 and also by scanning the morphology descriptors in columns (7) and (8) of 
Table~\ref{t_line}.  We note that, among the earlier Herbig Be stars, the 
phenomenon we have labelled depolarisation is the norm for stars 
presenting a clear line effect.  We remind this is the behaviour that is 
also 
seen in classical Be stars.  In contrast to this, there is a marked 
predominance of more rounded ``loops'' in the $(Q,U)$ plane plots for the 
HAe stars (seen in only one HBe star, the B9 object HD 58647).  In
\%Pol and PA as functions of wavelength, these loops correspond to
antisymmetric flips.  It was argued earlier that such behaviour may point 
to 
intrinsic \ha\ line polarisation that, in turn, most likely demands a 
compact 
source of \ha\ emission in these objects.  The fact that \ratpol\ tends to 
be 
noticeably less than one for these objects (Fig.~\ref{f_trend}) is 
consistent 
with this idea since the interior polarised line emission could well
be narrower than the brighter exterior unpolarised component.  

Among the HAe stars, and where the intensity profile includes a P~Cygni 
blueshifted absorption component, the above-mentioned loops can be 
accompanied by a second $(Q,U)$-plane excursion on the opposite side of 
the 
continuum knot (i.e. at a relative angle of 90$^o$ in the plane of the 
sky).  
Examples of this are AB~Aur and MWC 789.  In both cases, the 
``loop'' is due to the polarisation change across the central/redshifted 
line 
emission, while the second orthogonal ``excursion'' is associated with the 
blueshifted absorption.  Similar excursions may be present on their own in 
T~Ori and CQ~Tau where the polarisation change is associated with 
significant 
redshifted absorption.

From the above we find two limiting patterns of behaviour: (i) in the 
early B group (B0 -- B2) broad polarisation changes are observed across 
\ha\ that are consistent with depolarisation; (ii) in the later A stars (A2 -- F), 
the line effects associated with \ha\ emission tend to be narrower 
than the Stokes I profile, and loops are seen in the $(Q,U)$ plane. In between these 
two extremes we find transitional objects for which we have too few 
examples to be able to derive a general impression. 
It has to be regretted that later B spectral types remain poorly
sampled as it would be interesting to see how the transition from 
early B-type to the A-type \ha\ polarisation characteristics is
accomplished. 
Whether the detected difference between the HBe and the HAe stars is due 
to different physics in the HRD or instead due to some  
observational bias effect is now open for discussion.

\section{Summary \& Discussion}
\label{s_disc}

Altogether we have found that 16 out of 23 Herbig Ae/Be stars 
show a change in linear polarisation across \ha. Data on 19 objects
have been presented here, and are supplemented by 4 more from OD99.
Summarised:
 
\begin{itemize}
\item{} For the Herbig Be stars: 7 out of 12 reveal a line effect that is 
      consistent with a depolarisation effect caused by a flattened 
      structure in analogy to the presence of discs around classical Be 
      stars. 
 
\item{} For the Herbig Ae stars: 9 out of 11 show a line polarisation 
      effect characterised by a ``loop'' in $(Q,U)$ space, which suggests 
      compact \ha\ emission that is itself polarised by a rotating disc-like 
      medium. 

\end{itemize}

We have found that the changes detected in the Herbig Be stars are
like those observed in classical Be stars. It is worth considering whether
this 7/12 (or $\sim$58 percent) detection rate is also as high as found among classical
Be stars.  Before making such a comparison, it is necessary to establish
the typical sensitivity limit of our spectropolarimetry. This is most 
easily derived by assessing the radii of the continuum knots in the $(Q,U)$
plane plots: typically these are large enough that excursions out of them
that extend no further than 0.2\% are too unclear to claim as definite
detections.  Using this as the detection threshold for polarisation  
changes across \ha, we can turn for a comparison to the 
large classical Be sample of Poeckert \& Marlborough (1976).  They tabulated
the change in line polarisation percentage across \ha\ for 48 classical
Be stars (their Table 2). In 26 of these stars, the polarisation change was
greater than our sensitivity limit of 0.2\% . This amounts to a 54 percent 
detection rate to be compared with $\sim$58 percent here.  These percentages 
are strikingly alike.  The similarity in both spectropolarimetric behaviour 
and detection rate is a strong hint that Herbig Be stars are embedded in 
electron-scattering circumstellar discs of similar characteristics to
those associated with classical Be stars.  This allows the interpretation
of the non-detections to be examples of Herbig stars with their discs too 
face-on to yield measurable line depolarisations.

The predominance of $(Q,U)$ loops in the Herbig Ae star spectropolarimetry
suggests that compact \ha\ emission that is itself scattered and polarised
by a rotating, non-spherical medium is the norm.  Prior examples of this 
spectropolarimetric behaviour and discussion of its origin may be found in studies of 
the peculiar Be star $\gamma$~Cas (Poeckert \& Marlborough 1977) and 
the Wolf-Rayet binary EZ~CMa (Schulte-Ladbeck et al. 1990).  That the frequency 
of effects among the Herbig Ae stars is possibly higher than among the Herbig Be stars 
might be expected as
the PA rotation can remain visible over a wider range of viewing angles.
In the case of AB~Aur, Pontefract et al (2000) have suggested 
that the spectropolarimetric evidence of a rotating, flattened 
structure surrounding a compact \ha\ source might be due to 
magnetically-channelled accretion.  Catala et al. (1999) also invoked this
possibility in the face of a detection of time-variable
redshifted absorption in He{\sc i} lines in AB~Aur.  Our data include two
instances of stars showing significant redshifted absorption 
at \ha\ (T~Ori and CQ~Tau), and many more instances of $(Q,U)$ loops. 
(Note that infall features at \ha\ were previously found by e.g. 
Finkenzeller \& Mundt 1984; de Winter et al. 1999). 
Independently, the two lines of evidence would
not be such a strong indication of magnetically-channelled accretion -- but
together they are rather more compelling.  To pursue this issue  
further, an interesting next step would be to compare the linear 
spectropolarimetric characteristics for the Herbig stars with those in the 
lower mass T~Tauri stars, a class of 
young objects for which the magnetically-channelled accretion model is
better established (Edwards et al. 1994; Hartmann 1999).  
So far only circular spectropolarimetry has been published (Johns-Krull, Valenti 
\& Koresko 1999) for T~Tauri stars.  

We now consider why there is the observed difference between the Herbig Be 
and HAe groups, and what it might mean.  One point of view could be that
there is a physical transition region in the Hertzsprung-Russell Diagram
where the geometry changes, as possibly the mode of presumed accretion changes.
While the Herbig Ae stars might experience magnetically-channelled accretion 
resembling that associated with T~Tauri stars, the Herbig Be star accretion flow may 
not be disrupted at a magnetospheric radius. The possibility that this might be 
so inspired the exploration of a radiation-driven disc wind concept for Herbig Be 
stars and embedded Becklin-Neugebauer objects presented by Drew, Proga \& Stone (1998). 
Before the option of a real transition can be accepted it is important 
to recognise that the difference may instead simply be one of observational 
contrast.

Crudely speaking, spectral type maps onto effective temperature and, indeed, 
luminosity -- if differences in evolutionary age are taken to be a 
secondary influence in our sample.  Assuming all environmental
factors such as the geometry and quantity of circumstellar matter
around all Herbig Ae/Be stars to be the same, the mere facts of the
higher temperatures and luminosities of the HBe stars imply greater 
extension of the ionised regions around them, and of course larger 
dust-free cavities.  
In accord with this, \ha\ equivalent width is indeed greater among 
the HBe stars than among the HAe stars.  In the presence 
of reduced \ha\ line emission and opacity for the HAe stars it is
easier to detect the interior component that we require should exist 
inside an axisymmetric rotating structure. 
The failure to detect this at \ha\ in the early Be stars could then be 
nothing more than a masking of it by the sheer size of the total 
line-emitting volume and its optical depth.  

We have counted MWC~361 (see Fig.~\ref{f_hbe}) as an instance of broad line 
depolarisation.  This may not be the entire story.  The data for this object 
show a relatively narrow PA rotation in the red wing of \ha, on top of what
otherwise is a well-defined wide depolarisation.  This may hint at some 
interior \ha\ polarisation reminiscent of that seen among the HAe stars.  
Significantly, perhaps, the \ha\ equivalent width measured for MWC~361 (-63\AA) 
is lower than for most of the other early HBe stars showing simple 
depolarisations.
A difficulty may however be presented by the case of MWC~166 (HD~53367).
This early HBe star also has a relatively small \ha\ equivalent width ($-$2.8\AA\ 
rather and $-$14\AA\ in OD99), but no evidence of 
a compact \ha\ component. In other words, spectropolarimetrically it does 
not look like an HAe star.
The counter to this is to argue that a reclassification of this object is 
appropriate. Evidence to support this can be derived from the relative 
weakness of MWC 166's infrared continuum excess: Hillenbrand et al. (1992) 
classify it as a Class III source because it is so weak. 
These Class III objects are hard to distinguish from more mature
classical Be stars.  
Perhaps MWC~361 is the seemingly peculiar case that ultimately betrays the geometric 
parallels between the environments of HAe and HBe stars.  
However, caution
must be exercised in view of the potential impact of the known binarity
of MWC~361 (Millan-Gabet, Schloerb \& Traub 2001).  

There is a test that might distinguish between the two options discussed 
above. Spectropolarimetry could be performed across less opaque, higher 
excitation emission lines that can be expected to arise 
in the same location as the compact \ha\ component, if it exists in the HBes. If the 
extension and optical depth of \ha\ emission indeed masks the evidence 
of a compact source in the HBes, e.g. He~{\sc i} or higher Balmer line 
observations could reveal angle rotations ($Q,U$ loops) in both the 
HBe and the HAe stars. If however, angle rotations were not observed 
in the HBe population, while appearing among the HAe stars, then 
the balance of the argument would shift in favour of HBe environments 
being physically different from those of HAe stars on the smallest scales.

As we have found that flattened structures around
Herbig Ae/Be stars are common in our data, we wonder how this 
compares with the existing literature on this matter.
We have therefore checked if the spectropolarimetric behaviour of 
our HAe sample correlates at all with the so-called UXOr phenomenon 
(named after UX~Ori). The UXOr phenomenon is believed, by some authors, to be  
associated with dusty clouds orbiting the young star in a close 
to edge-on disc-like structure (see e.g. Natta et al. 1997, but 
see Herbst \& Shevchenko 1999 for a conflicting view). Six of the 
HAe stars from our sample appear in the HAe compilation 
by Natta et al., but only three of these undergo photometric changes 
larger than 1 mag in the V band characteristic for UXOrs\footnote{This is a
very loose description of the UXOr behaviour. Strictly speaking, the star
should undergo a colour reversal (blueing effect) which is characteristic 
for UX~Ori itself}. 
As some of the stars that do not show UXOr behaviour do show loops 
in $(Q,U)$ space, this indicates that the UXOr and spectropolarimetric 
behaviour have different physical origins:  
the UXOr behaviour may well be linked with large scale dusty clouds, 
while the spectropolarimetry yields insights on the structure of the 
accretion flow much closer to the star.

Recently, Millan-Gabet et al. (2001) conducted an interferometric 
study at near-infrared wavelengths and they concluded that accretion disc 
models can be ruled out in most cases. Instead it was claimed that spherical 
models reproduce the visibility data much better. Note the seeming 
contradiction with the findings of the spectropolarimetry. But what are the 
spatial scales involved? The Herbig Ae stars in Sect.~\ref{s_hae} of this 
study show loops in $(Q,U)$ space, associated with PA changes measured to be as 
narrow as $\sim$ 10 \ang. This wavelength width translates into a velocity in the  
rotating disc of about 225 \kms. Assuming a stellar mass of $\sim$ 2 \msun, we 
find that this rotation speed corresponds to spatial scales of only $\sim$ 7 stellar 
radii, or equivalently 0.07 AU. This is to be compared to the 
scales of 0.5 -- 5 AU that the interferometry of Millan-Gabet et al. 
can resolve. 

We conclude that the spectropolarimetry indeed probes the smallest-scale 
structures of the circumstellar medium, which are found to be 
flattened.

\paragraph*{\it Acknowledgements}

We thank Matthew Pontefract for his participation in obtaining 
the observations, and the referee Rens Waters for constructive comments.
The allocation of time on the William Herschel Telescope was awarded 
by PATT, the United Kingdom allocation panel.  JSV is funded by the 
Particle Physics and Astronomy Research Council of the United Kingdom.  
The data analysis facilities are provided by the Starlink Project, which 
is run by CCLRC on behalf of PPARC. This research has made use of the
{\sc simbad} database, operated at CDS, Strasbourg, France.

\end{document}